\documentclass[final,3p,times,twocolumn]{elsarticle}
\usepackage{graphicx}
\usepackage{amssymb}
\usepackage[version=4]{mhchem}
\usepackage[dvipsnames]{xcolor}
\usepackage{siunitx}
\usepackage{lmodern}
\usepackage[hidelinks]{hyperref}
\usepackage[english]{babel}
\usepackage[modulo]{lineno}
\usepackage{soul}
\usepackage{tikz}
\usepackage{subcaption}
\newcommand{\dd}{\mbox{d}}
\newcommand{\norm}[2][2]{\left\lVert#2\right\rVert_#1}
\journal{Chemical Engineering Science}
\labelformat{figure}{#1}
\newcommand{\der}[2]{\frac{\mathrm{d}#1}{\mathrm{d}#2}}
\newcommand{\parder}[2]{\frac{\partial#1}{\partial#2}}

\begin{document}
\begin{frontmatter}
  \title{Can we live Danckwerts' dream? Mixing Analysis in a Baffled Stirred Tank Reactor Based on 4D-Particle Tracking Experiments}

  \author[1]{Eike~Steuwe\corref{cor1}}
  \ead{eike.steuwe@haw-hamburg.de}
  \cortext[cor1]{Corresponding author}
  \author[2]{Thanh Tung Thai}
  \author[3]{Christian Weiland}
  \author[2]{Anna Kl\"unker}
  \author[1]{Jan Hendrick Nissen}
  \author[3]{Michael Schl\"uter}
  \author[2]{Kathrin Padberg-Gehle}
  \author[1]{Alexandra~von~Kameke\corref{cor1}}
  \ead{alexandra.vonkameke@haw-hamburg.de}

  \address[1]{Faculty of Sustainable Engineering, Hamburg University of Applied Science}
  \address[2]{Institute of Mathematics and its Didactics, Leuphana University L\"uneburg}
  \address[3]{Institute of Multiphase Flows, Hamburg University of Technology}

  % ===================================================================
  % ===================================================================
  
  \begin{abstract}
    We present an experimental investigation of mixing dynamics within a laboratory-scale 3-liter stirred tank reactor (STR) equipped with two Rushton turbines and three baffles. Using time-resolved, four-dimensional particle tracking velocimetry, we successfully capture trajectories of up to 40,000 tracer particles in the full reactor volume despite obstructions by stirrer and baffles, providing unprecedented time-resolved flow and mixing information. From these Lagrangian data, we analyze velocities, accelerations, and spatial dispersion, revealing anisotropic mixing. By utilizing novel network-based analysis methods on the experimental particle trajectories, we identify coherent fluid compartments that exhibit strong internal mixing but weak exchange with neighboring compartments. We uncover five distinct compartments acting as transport barriers, which have a high impact on substrate distribution in chemical and biochemical processes. Our approach thus realizes and extends early thought experiments from Danckwerts and Levenspiel by providing detailed insight into the behavior of single fluid parcels and Lagrangian mixing withing chemical and biochemical reactors, offering a valuable approach for evaluation and optimization of chemical and biochemical processes. The trajectory data are made freely available to serve as an experimental reference for further research.
  \end{abstract}

  % ===================================================================
  
  %\begin{graphicalabstract}
  %\includegraphics[width=\textwidth]{figures/grabs.pdf}
  %\end{graphicalabstract}

  % ===================================================================
  % ===================================================================
  
 % \begin{highlights}
 %  \item Following \num{40000} fluid parcels through a stirred tank reactor using 4D-PTV
 %  \item Analysis of the encounters of fluid elements using novel network analysis methods
 %  \item Five coherent compartments hindering macromixing are identified
 %  \item Flow patterns, mixing, dispersion and velocity statistics in full 3D tank volume
 %  \item Experimental data exhibits a high spatial accuracy of 20 µm 

 %  \end{highlights}

\end{frontmatter}
\newpage

% ===================================================================
% ===================================================================

\section{Introduction}\label{sec:intro}
For many liquid conversions, mixing behavior plays an important role, since an inhomogeneous substrate distribution can direct chemical reactions towards different sometimes undisired intermediates or products~\cite{chanfrau_influence_2018,paul_handbook_2010,bourne_mixing_2003}. Zones in which the fluid is mainly contained over a long period of time with little to no interaction with neighboring areas, called compartments~\cite{banisch_understanding_2017}, can inhibit mixing processes. In bioreactors in particular, a homogeneous substrate distribution is usually necessary since concentration gradients can cause microorganisms to change their metabolic pathways. This can result in the formation of unwanted by-products and a reduction in yield and selectivity.

In the 1950s, Danckwerts made significant progress in characterizing mixing within non-ideal chemical reactors by introducing the residence time distribution~\cite{danckwerts_continuous_1953}.
However, as Danckwerts himself notes, his residence time distribution theory is ``seldom applicable to practical problems except as a rough guide'' since ``firstly, \ldots the chance of a molecule reacting depends on its path through the reactor, as well as its residence-time. Secondly, if the reaction is of order other than first, the chance of a given molecule reacting depends on the molecules which it encounters in its passage through the reactor.''~\cite[pp.~10--11]{danckwerts_continuous_1953}
Obtaining this Lagrangian information about the reacting molecules or even only the full Eulerian character of a three-dimensional flow, ``requires experimental measurements of greater accuracy than [they] were able to obtain''~\cite{kristmanson_studies_1961} at the time. It was really unimaginable to obtain such detailed flow and mixing information back then, as also the quote by Levenspiel~\cite[pp.~243--244]{levenspiel_chemical_1962} reveals: ``To predict the exact behavior of a vessel as a chemical reactor we must know what is happening in it\ldots.
And how can this be found? There is only one way, and that is to tag and follow each and every molecule as it passes through the
vessel\ldots Though fine in principle, the attendant complexities make it impractical to use this approach.''

Since then luckily, flow measurement technology has greatly advanced to the point that high-speed, high-resolution cameras and high-speed light sources allow for capturing particle images from which the three-dimensional flow fields can be reconstructed~\cite{kameke_experimental_2026}. Recent advances in these measurement techniques have made it possible to track large numbers of individual microscopic particles moving passively with the flow, bringing the realization of Danckwerts' dream within reach (particle tracking velocimetry, PTV)~\cite{schanz_shakethebox_2016, hofmann_lagrangian_2022}. However, up to now measurements have been restricted to subvolumes of reactors, since obstruction due to internals like stirrer and baffles has hindered optical accessibility~\cite{hofmann_lagrangian_2022,kuschel_validation_2021}. This led to interrupted trajectory data, ruling out the continuous monitoring of particle-particle encounters or the single particle residence times in subvolumes of the full reactor. In this work, we report a 4D-PTV measurement of up to \num{40000}~particles in the full volume of a \qty{3}{\liter} stirred tank reactor, operated in batch mode. To the best of our knowledge, this is the first time that Danckwerts' idea of deriving the Lagrangian information of the flow in a full reactor has been experimentally realized. From these experimental Lagrangian data various different results can be derived and for further wide usage, the trajectory data are made freely available (\url{https://doi.org/10.15480/882.17469}).

Stirred tank reactors are often key components in chemical and biochemical processes due to their flexibility and ease of adjustment~\cite{reschetilowski_handbook_2026}.
Additionally, the ideal stirred tank, that assumes that the whole volume is completely mixed, serves as an important model system for the design of reaction paths and reaction times~\cite{kraume_Transportvorgaenge_2012}. Further, the concept of an ideal stirred tank is used in compartment modeling~\cite{levenspiel_chemical_1999}. In practice, however, the assumption of an ideal, fully mixed reactor volume is often invalid, since it depends on the mixing dynamics, which are usually considerably slower than the reaction times. The mixing dynamics in turn depend on the scale-up level and thus the assumption of ideal mixing needs to be verified for each setting. In the batch setting, the concept of residence times is not directly applicable. Usually for batch reactors in industrial settings the mixing time is measured instead of the residence time~\cite{weiland_introduction_2023}. Elaborated methods for its determination have been developed during the last years~\cite{fitschen_novel_2021}. However, these experiments oftentimes reveal that mixing efficiency in compartments might be higher or lower than in the remaining regions of the reactor and that these mixing inhomogeneities vary considerably for each experimental trial.
Here we present a novel Lagrangian analysis method that uses the 4D-PTV trajectories to obtain these compartments. To this end, the analysis closely follows the previously introduced ideas of Danckwerts and Levenspiel to evaluate all the contacts of a molecule on its path through the reactor. Of course, we cannot follow every molecule, but we can follow each of the \num{40000}~flow-following particles in its place. These can be interpreted as fluid parcels or molecules, and their interaction can be mathematically formulated by means of adjacency matrices. These form the basis of elaborated network analysis methods~\cite{schneide_lagrangian_2019} from which compartments can be derived. We show that these new methods have the potential to predict mixing, dissolution and reaction processes.

The current paper is organized as follows: In Section~\ref{sec:methods} the experimental details, data processing and the network-based analysis method are described. In Section~\ref{sec:results} the resulting Lagrangian data and statistics are discussed and the radial and axial dispersion is evaluated from the trajectory data. Also, it is shown how Eulerian quantities can be derived from these trajectories and how they relate to the Lagrangian data. Finally, the results of the network-based analysis of the trajectory data are presented which reveal compartments that govern the mixing in the stirred tank.

% ===================================================================
% ===================================================================

\section{Material and methods}\label{sec:methods}

The following presents the experimental setup used to obtain the trajectory data. This is then followed by a description of the methods used for data processing and analysis.

\subsection{Experimental setup}
\begin{figure*}[ht!]
  \centering
  \includegraphics[width=.7\textwidth]{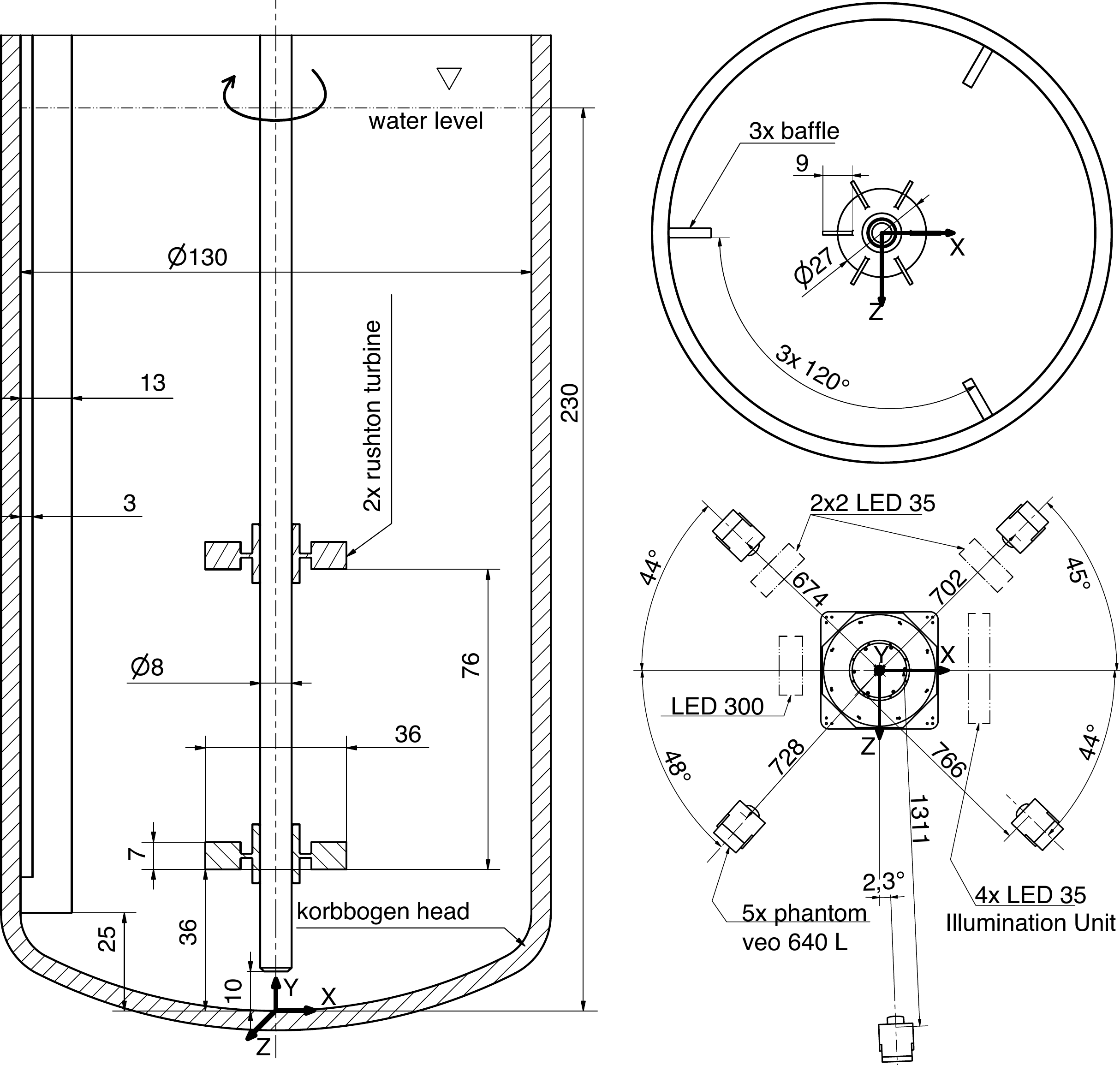}
  \caption{Technical drawing of the STR used and positioning of the cameras surrounding the STR during the measurement. The fluid is stirred with two Rushton turbines and diverted by three equidistant flow breaking baffles. The camera at farthest away is tilted downwards by \qty{11}{\degree}; all other cameras are tilted downwards by \qty{22}{\degree}. Each camera's field of view covers the entire STR.\@ All dimensions are given in millimeters.}\label{fig:methods_expsetup}
\end{figure*}

The experiments were carried out in a three liter stirred tank reactor (STR) with three baffles and two Rushton turbines at atmospheric pressure. The STR is a handmade glass vessel with an inner diameter of $d_\text{R}=\qty{130}{\milli\meter}$ and a Korbbogen head bottom. Three 3D-printed baffles, made of black polylactid acid (PLA), were arranged at 120 degree intervals along the reactor wall. The vessel was equipped with two Rushton turbines powered by a MINISTAR 80 control (IKA-Werke GmbH \& Co. KG). All relevant dimensions of the lab-scaled STR are included in Figure~\ref{fig:methods_expsetup}. STRs with similar dimensions have been previously investigated experimentally~\cite{fitschen_novel_2021,hofmann_lagrangian_2022} and computationally~\cite{weiland_computational_2023} and are also used as a laboratory standard in industry, e.g.\ as scale down models for bioreactors~\cite{piontek_modeling_2026}.

In order to compensate for different refractive indices and improve visual access through the curved vessel wall, the STR was placed in an octagonal-shaped prism filled with bidistilled water. The prism had an edge length of \qty{165}{\milli\meter} and was made of \qty{5}{\milli\meter} thick polycarbonate. Five synchronized cameras (Phantom VEO 640L) were positioned around the STR as shown in Figure~\ref{fig:methods_expsetup} so that, despite obstruction of the built-in components, every point in the investigated volume was visible to at least two cameras. Each camera was positioned at an elevated angle and tilted downward, providing different lines of sight of the investigated volume. Four cameras were equipped with ZEISS Milvus 1.4/50 lenses, while the one placed furthest away was equipped with the ZEISS Milvus 2/100M. On each lense, a low-pass filter was mounted that cuts off light with a wavelength smaller than \qty{550}{\nm} and a Scheimpflug was positioned in between the cameras and the objective lenses. For illumination, blue high-power LEDs (one LaVision LED300 and eight LaVision LED35s), with wavelengths of \qty{445}{\nm}, were synchronized with the frame rate of the cameras.
The STR was filled with $V_\text{R}=\qty{3}{\liter}$ bidistilled water at room temperature (\qty{22}{\celsius}), resulting in a water level height of $h_\text{R}=\qty{230}{\milli\meter}$. The water was seeded with different amounts of `Fluorescent Red Polyethylene Microspheres' (Cospheric LLC).
The number of seeded particles range from approximately \qty{350}{particles} for volume self-calibration purposes to approximately \num{40000}~particles for the measurements reported here. The used particles have a density $\rho_\text{P}$ of \qty{995}{\kg\meter\tothe{-3}} and a diameter $d_{\text{P}}$ of \qtyrange{45}{53}{\micro\m}. According to Ouellette et al.~\cite{ouellette_transport_2008}, the Stokes number can be calculated as
\begin{equation*}
  St = \dfrac{2 \cdot \rho_{\text{P}} \cdot r_{\text{P}}^2}{9 \cdot \rho_\text{f} \cdot d_{\text{S}}^2} Re
\end{equation*}
where $Re$ is defined as
\begin{equation*}
  Re = \dfrac{\omega \cdot d_\text{S}^2 \cdot \rho_\text{f}}{2 \eta_\text{f}}.
\end{equation*}
This results in
\begin{equation*}
  St = \dfrac{\omega \cdot \rho_{\text{P}} \cdot d_{\text{P}}^2}{36 \eta_\text{f}}.
\end{equation*}
The stirrer speed was set to \qty{252}{rpm}, resulting in a Stokes number of $St < 10^{-2}$, where $\omega$ and $d_\text{S}$ denote the angular velocity and diameter of the stirrer, $\rho_{\text{P}}$, $r_{\text{P}}$ and $d_{\text{P}}$ denote the density, radius and diameter of the particle, while $\eta_\text{f}$ and $\rho_\text{f}$ denotes the dynamic viscosity and density of the surrounding fluid. Although the particle trajectories of the tracer particles used might still slightly diverge from perfect tracer trajectories at distinct dynamic areas of the flow field~\cite{ouellette_transport_2008}, a Stokes number significantly smaller than one is still the simplest measure to assure that the particles follow the flow with as little a lag as possible~\cite{lau_effect_2016}.
The impeller Reynolds number \[Re=\dfrac{\omega \cdot d_\text{S}^2 \cdot \rho_\text{f}}{2\pi \eta_\text{f}}\] with a value of $Re\approx \num{5700}$ indicates that the STR was operated in a transitional flow regime, which can usually be found for $\num{10}\le Re \le \num{10000}$ in stirred tanks~\cite{doran_bioprocess_2013}. Here, $d_\text{R}$ denotes the diameter of the stirrer.

\subsection{Calibration}
A customized calibration target was created using a black anodized, rectangular aluminum plate with rounded corners. This plate was engraved with a pointed pattern to ensure precise calibration of nearly the full STR volume. The pinhole fit implemented in DaVis 11.0 software (LaVision GmbH, Germany) was then used to determine the camera positions in three-dimensional space relative to the calibration target's position. Therefore, the internal components of the STR were removed, and the STR was filled with bidistilled water. The calibration target was placed in the STR and tilted at arbitrary angles while images were taken with all cameras simultaneously illuminated by ambient light. Afterward the target was removed and approximately 350~tracer particles were added. The fluid was manually agitated at an arbitrary speed, and images of the particles, illuminated by the high-power LEDs, were captured at a rate of \qty{100}{\Hz}.
These images were used to refine the calibration function to subpixel precision by iteratively utilizing the volume self-calibration implemented in DaVis. The pinhole fit was translated into two-dimensional third-order polynomials $f:(x,y,z)\longmapsto (x'_i,y'_i)$ for each camera, which project the world coordinates $(x,y,z)$ to camera image positions $(x'_i,y'_i)$, here $i$ denotes the $i^\text{th}$ camera.

\subsection{Measurement and processing}\label{sec:processing}
\begin{figure*}[ht]
  \centering
  \includegraphics[width=.9\textwidth]{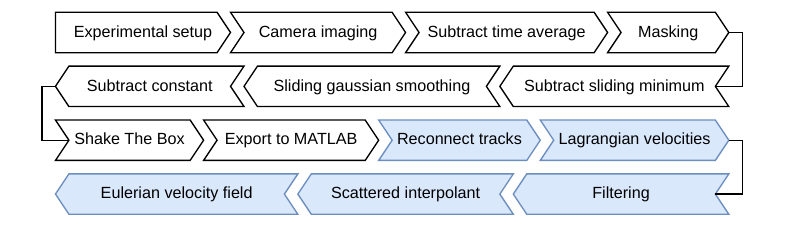}
  \caption{Data processing flowchart. Processes performed in the DaVis software have a white background and those in MATLAB are colored blue.}\label{fig:methods_processing_flowchart}
  \color{black}
\end{figure*}

\begin{figure*}[ht]
  \centering
  \includegraphics[width=\textwidth]{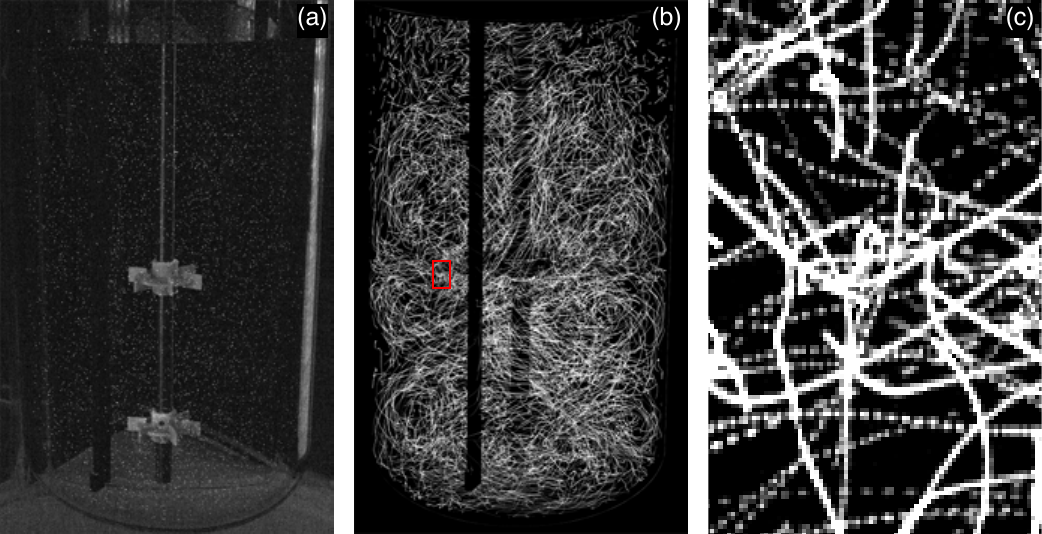}
  \caption{Images of the stirred tank reactor seeded with approximately \num{5000}~fluorescent particles.\@(a) Single raw image.\@(b) Sum of \num{120} preprocessed images, covering one stirrer rotation.\@(c) Close-up view of the red-highlighted region in (b).}\label{fig:methods_raw_processed_images}
  \color{black}
\end{figure*}

In a first step, approximately \num{1200} particles were added, and the baffles and stirrer were installed to take the measurements. A total of \num{2950} images, each measuring \num{2560}~$\cdot$~\num{1600}~pixels, were captured at a frame rate of \qty{500}{\Hz} and an illumination time of \qty{1}{\ms} using high-power LEDs. The measurement was repeated for approximately \num{5000}, \num{10000}, \num{20000}, and \num{40000} particles, with the latter being presented and discussed in detail here. In this case \num{2818}~images could be processed. The data processing is summarized in Figure~\ref{fig:methods_processing_flowchart}.
The images were preprocessed using the DaVis software. First, the time-averaged image was subtracted from each camera's image, respectively. Then, the image was tailored to the recorded STR outline using a polygonal mask, and a spatial sliding minimum filter with a size of $5 \cdot 5$~pixels, was applied to normalize the image brightness to local illumination conditions.
Finally, a Gaussian filter with a standard deviation of one pixel was applied, after which the intensity of the entire image was reduced by 20~counts to cut off the noise. These preprocessed images were used to generate an optical transfer function (OTF) within the DaVis software, representing particle images as elliptical Gaussian blobs for different subvolumes. Figure~\ref{fig:methods_raw_processed_images} shows the effect of image preprocessing. The images presented here are taken from a measurement with a low particle count of \num{5000} such that the particle trajectories are readily recognizable. Figure~\ref{fig:methods_raw_processed_images}a shows a raw image, where the stirrers and the background are visible, as are the reflections of the high-power LEDs on the glass surface, the liquid surface, and the stirrer shaft, which disturb the image.
Figure~\ref{fig:methods_raw_processed_images}b shows preprocessed images summed over one stirrer rotation, making the particle trajectories visible. Image preprocessing results in the cancellation of reflections and a reduction in overall noise. The close-up view in Figure~\ref{fig:methods_raw_processed_images}c shows that the particles are visible as approximately Gaussian blobs with a mean diameter of two to three pixels after image preprocessing. 

The $N_{\text{P}}$ particle trajectories $\boldsymbol{x}_{P}(t) \in \mathbb{R}^3$, $P = 1, \dots, N_{\text{P}}$ are obtained at discrete times $t_k \in \mathbb{T}\subset{\mathbb{R}}$, $k = 1, \dots, N_\text{T}$, where $\mathbb{T}$ is the set of the discrete times and $N_\text{T}$ is the cardinality of $\mathbb{T}$, from the preprocessed images using the Variable Time-Step Shake-The-Box (\mbox{VT-STB}) method implemented in the DaVis software~\cite{schanz_shakethebox_2016,schroder_3d_2023,schanz_shakethebox_2021}, which utilizes the calibration function to project the particle positions from the camera images back into three-dimensional space. The \mbox{VT-STB} is a recursive algorithm that reconstructs particle trajectories using variable step sizes throughout the recorded images. Particles found in one run are subtracted from the image for the next run using the OTF, meaning they are not detected in subsequent runs again. Varying the step size allows the algorithm to detect and track both fast- and slow-moving particles within the investigated volume making the recording of highly dynamic flows possible.
The resulting particle trajectories were exported from DaVis as machine- and human-readable\ .dat files and imported into MATLAB.\@

The entire data set had to be shifted and rotated due to calibration bias, in order to align it with the coordinate system shown in Figure~\ref{fig:methods_expsetup}. The correct position was determined by visually inspecting the projection of the particles onto the three principal coordinate planes.
To further improve the quality of the data set, a custom-made MATLAB algorithm was used to reconnect interrupted particle trajectories. This algorithm extrapolates the start and end of interrupted trajectories linearly for 15~time steps. Then, the root mean square (RMS) is calculated for each pair of starting and ending trajectories at every time step. Not only the extrapolated points are considered but also the first and last 15 detected positions, accounting for occasional duplicate particle detections by the \mbox{VT-STB} algorithm. Starting with the lowest RMS, the trajectory pairs are matched successively until the RMS of the remaining trajectory pairs is higher than a threshold. An adequate threshold can be calculated based on the uninterrupted trajectories. Therefore, the positions of two consecutive time steps of a track are used to calculate the subsequent position by linear extrapolation.
The RMS of the difference between this extrapolation and the actual position is determined for all uninterrupted trajectories. The final threshold used for the matching step is the 99\textsuperscript{th} percentile of the RMS values multiplied by a factor of five. The matched trajectories are reconnected using a third-order polynomial fitted to the last five points of the trajectory ends, respectively. The polynomial is then evaluated at the missing time steps.
Through visual inspection of the trajectory differences before and after reconnection, no falsely reconnected trajectories were found. After reconnecting, the trajectories are each smoothed using the MATLAB implementation of a Savitzky-Golay filter~\cite{savitzky_smoothing_1964} with a kernel size of 15~time steps and an order of three. These filter parameters turned out to be a good choice to cut out noise induced by the \mbox{VT-STB} algorithm tracking subpixel scintillations in the particle images while keeping the whole trajectory in its original position and shape.
\begin{figure*}[ht!]
  \centering
  \includegraphics[width=.99\textwidth]{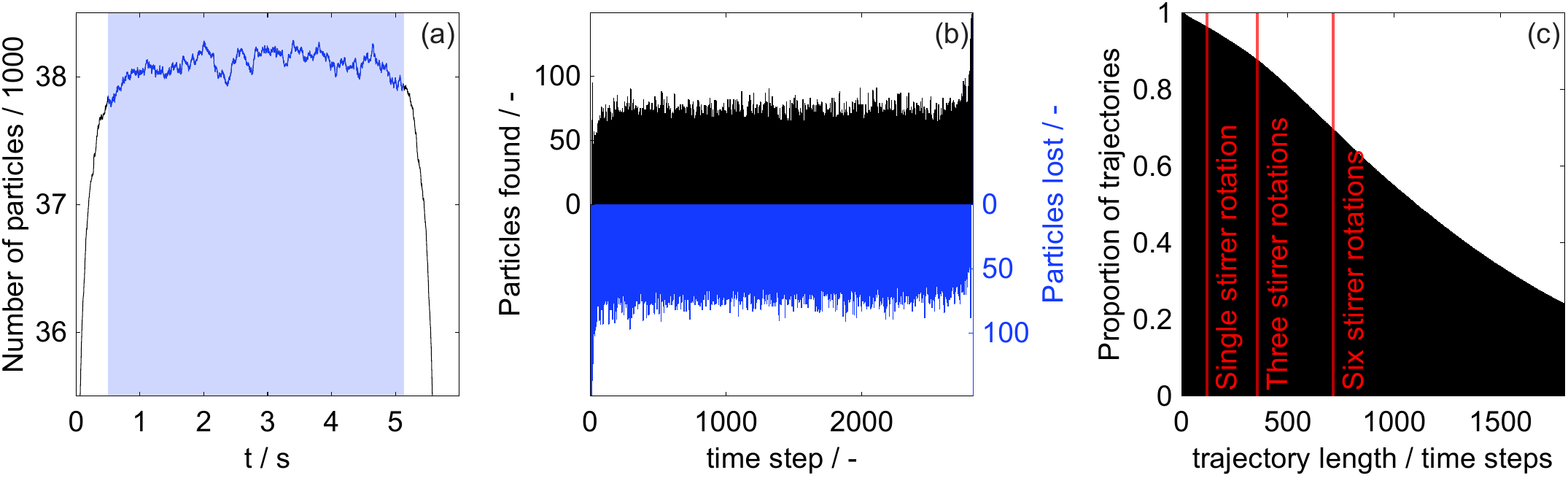}
  \caption{Particle trajectory statistics.\@(a) Number of tracked particles versus time. Particles outside the blue-highlighted time interval are discarded.\@(b) Number of particles lost and found within one time step.\@(c) Cumulative proportion of trajectories of the length as indicated on the horizontal axis. }
  \color{black}
  \label{fig:methods_trajectories_statictics}
\end{figure*}
As shown in Figure~\ref{fig:methods_trajectories_statictics}a, due to the nature of the \mbox{VT-STB} algorithm, fewer trajectories are found at the beginning and end of the investigated time interval than on average for the other time steps.
After discarding the first and last \qty{0.5}{\second} of the data set, \qty[separate-uncertainty = true]{38100(95)}{trajectories} are found at every time step, where the latter constitutes the standard deviation.
Thus, the final considered time interval used for analysis of this measurement is reduced to approximately \qty{4.6}{\second}, equivalent to \num{19.3}~stirrer rotations.
Figure~\ref{fig:methods_trajectories_statictics}b is a bar diagram showing the number of particles lost and found at each time step. On average, both values are \qty{65}{trajectories}, equivalent to \qty{0.17}{\percent} of all trajectories. Figure~\ref{fig:methods_trajectories_statictics}c provides the proportion of trajectories exceeding the respective length in time steps plotted on the horizontal axis. The time intervals marked by the red vertical lines are discussed in detail in Section~\ref{sec:discussion_dispersion} and Section~\ref{sec:discussion_network}. For the Lagrangian analysis, only particles tracked for longer than the investigated time interval are usable. Figure~\ref{fig:methods_trajectories_statictics}c shows that, for the chosen time intervals, \qty{96.4}{\percent}, \qty{87.7}{\percent} and \qty{70.0}{\percent} of all trajectories are usable for the analysis of one, three, and six stirrer rotations, respectively.

The Lagrangian velocities $\boldsymbol{v}_{P}(t) = \der{\boldsymbol{x}_{P}(t)}{t}\in \mathbb{R}^3$, ${{P}} = 1, \dots, N_{\text{P}}$ for each particle are evaluated at discrete times $t_k \in \mathbb{T}$ by finite differencing along all particle trajectories in time, using a difference quotient of 4\textsuperscript{th}~order accuracy~\cite{fornberg_generation_1988}
\begin{equation*}
    \begin{split}
        \boldsymbol{v}_{{{P}}}(t_k)\approx
      \frac{1}{12\delta t} &\cdot (\boldsymbol{x}_{{{P}}}\left(t_{k-2}\right)-8\boldsymbol{x}_{{{P}}}\left(t_{k-1}\right)\\&+8\boldsymbol{x}_{{{P}}}\left(t_{k+1}\right)-\boldsymbol{x}_{{{P}}}\left(t_{k+2}\right)) \;,\\
      k&=3, \ldots, N_T-2 \; ,
    \end{split}
\end{equation*}
where $\delta t$ is the constant time step, i.e.\ $t_{k+1}=t_k+\delta t$.
As long as the spatial difference of consecutive positions of a particle is small relative to the spatial change of the flow and the time difference ${\delta t}$ is small relative to the temporal change of the flow, this approximation is valid.
Two representations of $\boldsymbol{v}_{{{P}}}(t)$ in different bases are used throughout the article, namely $\boldsymbol{v}^{\text{cart}}_{{{P}}}(t) = {\left( v_{{{P}},x}(t),v_{{{P}},y}(t),v_{{{P}},z}(t) \right)}$ in Cartesian coordinates and $\boldsymbol{v}^{\text{cyl}}_{{{P}}}(t) = {\left( v_{{{P}},\phi}(t),v_{{{P}},y}(t),v_{{{P}},r}(t) \right)}$ in cylindrical coordinates, where $v_{{{P}},\phi}(t) = - v_{{{P}},x} \cdot \sin(\phi) + v_{{{P}},z} \cdot \cos(\phi)$ and $v_{{{P}},r}(t) = v_{{{P}},x} \cdot \cos(\phi) + v_{{{P}},z} \cdot \sin(\phi)$.
From the Lagrangian particle velocity the Eulerian velocity $\boldsymbol{v}(\boldsymbol{x},t) = (v_x(\boldsymbol{x},t),v_y(\boldsymbol{x},t),v_z(\boldsymbol{x},t))$ at the particle positions $\boldsymbol{x}_{{{P}}}(t)$ can be derived by

\begin{equation*}
  \boldsymbol{v} (\boldsymbol{x}_{{{P}}}(t), t) = \boldsymbol{v}_{{{P}}}(t)
  \; .
\end{equation*}

Thus, the Eulerian velocity field at a specific time step $t_k$ can be obtained from the set of all particle velocities $\boldsymbol{v}_{{{P}}}(t_k),{{P}} = 1, \dots, N_{\text{P}}$. The Eulerian velocity is then only given at the positions of the particles and, therefore, at randomly distributed points within the investigated volume. To obtain the velocity between these points, the MATLAB built-in function \texttt{scatteredInterpolant} is used to interpolate between the given points using Delaunay triangulation~\cite{delaunay_sphere_1924}.
By evaluating the \texttt{scatteredInterpolant} at evenly distributed points, a Eulerian velocity field can be created for an evenly distributed cartesian grid. Then the Eulerian acceleration field and spatial velocity field gradients from which insightful measures such as the energy dissipation rate $\epsilon = 2 \cdot \boldsymbol{\mathsf{T}\mathsf{T}}\cdot\eta_\text{f}/\rho_\text{f}$ can be calculated from the square of the symmetric strain rate tensor

\begin{equation*}
\begin{split}
\boldsymbol{\mathsf{T}\mathsf{T}} = 
\left(\parder{v_x}{x}\right)^2
+\left( \parder{v_y}{y}\right)^2
+\left( \parder{v_z}{z}\right)^2 \\
+\frac{\left(\parder{v_x}{y}+\parder{v_y}{x}\right)^2
+\left(\parder{v_y}{z}+\parder{v_z}{y}\right)^2
+\left(\parder{v_z}{x}+\parder{v_x}{z}\right)^2}{2}
\end{split}
\end{equation*}
and the dynamic viscosity $\eta_\text{f}$ and density $\rho_\text{f}$ of the fluid under the measurement conditions~\cite{schroder_measurements_2022}. For the current study, the velocity data $v_x$, $v_y$, and $v_z$ are evaluated from the scattered interpolant on a grid of size $(131\times235\times131)$ with a spacing of $\approx$ \qty{1}{\milli\meter}, which oversamples the mean inter-particle distances of the PTV measurement of $N_{\text{P}}/V_R = \qty{0.0133}{\per\cubic\milli\meter}$ by a factor of approximately \num{4.2} in each spatial dimension and guarantees smoothness of the approximation of the velocity field and its derivatives also in highly dynamic flow regions.
Furthermore, the resulting velocity fields are slightly smoothed using the built-in MATLAB function \texttt{imgaussfilt3} with a standard deviation of $0.75$ and zero padding. To ensure that the velocity at the boundaries of the reactor goes to zero and is not misled by single erroneously detected ghost particles, dummy particles at the reactor boundary positions $\boldsymbol{x}_{b}$ are added to the other trajectories with $\boldsymbol{v}(\boldsymbol{x}_{b},t_k) = 0$ for all $t_k \in \mathbb{T}$ and $b\in\mathbb{B}\subset\mathbb{N}$, where $\mathbb{B}$ is the set indicating boundary positions, before generating the scattered interpolant functions from the particle trajectories. These dummy particles sit on a cartesian grid that is generated slightly larger than the vessel with \qty{60} grid points in each spatial dimension from the geometry .stl file of the STR.
In addition, the Eulerian velocities can also be evaluated in cylindrical coordinates by the relation $(v_\phi,v_y,v_r) = (-v_x \sin(\phi) + v_z \cos(\phi),v_y,v_x \cos(\phi) + v_z \sin(\phi))$.

% =========================================================
% =========================================================

\subsection{Network-based analysis of transport and mixing}
\label{sec:methods_network}
In order to evolve Danckwerts' early ideas to follow every molecule into a tractable mathematical description that allows us to study Lagrangian transport and mixing processes in reactors, a network-based approach is applied. It makes direct use of the measured Lagrangian particle trajectories and represents the interaction of particles over time in a network as originally proposed in~\cite{padberg-gehle_networkbased_2017} for geophysical flows. Following the derivations in~\cite{schneide_evolutionary_2022}, at each time instance $t_k$ an instantaneous binary adjacency matrix $\boldsymbol{A}(t_k) \in \{ 0, 1 \}^{N_{\text{P}}}$ is constructed with entries
\begin{equation*}
  a_{ij}(t_k) =
  \begin{cases}
    1 & \quad \text{if} \quad \norm{\boldsymbol{x}_i(t_k) - \boldsymbol{x}_j(t_k)} \leq \epsilon_r, \; i\neq j \\
    0 & \quad \text{otherwise.}
  \end{cases}
\end{equation*}
Thus, $a_{ij}(t_k)=1$ when the two different particles $i$ and $j$ have come $\epsilon_r$-close in time step $t_k$ and $a_{ij}(t_k)=0$ otherwise. Note that the diagonal $a_{ii}(t_k)=0$ for all $t_k \in \mathbb{T}$ and all $i=1,\ldots, N_{\text{P}}$. The threshold $\epsilon_r > 0$ defines an interaction radius. This can be chosen based on either physical considerations as in~\cite{schneide_evolutionary_2022} or to target a certain link density in the resulting network~\cite{donner_ambiguities_2010}. SciPy's \texttt{sparse\_distance\_matrix} implements such an $\epsilon_r$-neighborhood search based on KDTrees.

By summing up the instantaneous adjacency matrices $\boldsymbol{A}(t_k)$ for all time steps, the weighted network matrix $\boldsymbol{W}\in\mathbb{N}_0^{N_\text{P}\times N_\text{P}}$ with entries
\begin{equation*}
  {w}_{ij} = \sum_{k=1}^{N_\text{T}} a_{ij}(t_k)
\end{equation*}
is obtained. This symmetric matrix implements temporal information of how often a particle was within the $\epsilon_r$ range of every other particle, thus encoding the number of total $\epsilon_r$-close encounters over the evaluated time interval $\Delta t = [t_1, t_{N_\text{T}}]$. The entry $w_{ij}$ is therefore large if the trajectories of the two different particles $i$ and $j$ stay close for a long time or if the two particles repeatedly meet in terms of $\epsilon_r$-close encounters.

If these particles are viewed as fluid parcels, these will have plenty of time for their liquids to interchange and to react, if initially carrying different reactants. Thus, they will rapidly approach a similar level of reactant concentrations. We use this characteristic to define coherent volumes, here called compartments, that travel together later in the text. These compartments comprise particles that may mix within, but do not mix well with other particles outside these coherent compartments. Depending on their size, which can be controlled in the analysis framework (e.g.\ by the choice of $\epsilon_r$), these coherent compartments thus hinder the overall macro-mixing but allow for meso- and micro-mixing in the corresponding volume~\cite{fitschen_novel_2021,johnson_chemical_2003}.

If we want to detect those fluid parcels that mix fastest instead, we are not interested in repeated encounters of particles but rather in those particles that meet the most of all other particles. Such ``spreaders'' are rapidly moved from one part of the reactor to another and macro-mixing is largely enhanced. To study these fluid parcels, we define the general adjacency matrix $\boldsymbol{\bar{A}}\in \{ 0, 1 \}^{N_{\text{P}}}$ with entries
\begin{equation*}
  \bar{a}_{ij} =
  \begin{cases}
    1 & \quad \text{if} \quad w_{ij} > 0 \\
    0 & \quad \text{otherwise.}
  \end{cases}
\end{equation*}
This matrix implements whether two particles have come $\epsilon_r$-close over the respective time span at all, and thus the sum of all values in a row is $N_{\text{P}}-1$ whenever a particle has met all other particles.
In this sense, the unweighted node degree vector $\boldsymbol{\xi}\in\mathbb{N}_0^{N_\text{P}}$ with the entries
\begin{equation}
  \xi_{i} = \sum_{j=1}^{N_{\text{P}}} \bar{a}_{ij},
  \label{eq:unw_node_degree}
\end{equation}
is obtained from $\boldsymbol{\bar{A}}$. Here, $\xi_{i}$ encodes the number of different particles that a specific particle $i$ has encountered during the observed discrete time interval. The node degree value of each particle can be plotted, e.g., at its initial location at the beginning of the considered time interval $\Delta t = [t_1, t_{N_\text{T}}]$.
Regions in the reactor with high mixing efficiency correspond to high node degree values. Low node degree values indicate regions with poor mixing efficiency.
In~\cite{banisch_network_2019} a formal link to finite-time Lyapunov exponents is established, which is a heuristic Lagrangian concept frequently used for identifying transport barriers in fluids~\cite{haller_lagrangian_2015}.

Mathematically, the same can be done using $\boldsymbol{W}$, and the weighted node degree vector $\boldsymbol{\zeta}\in\mathbb{N}_0^{N_\text{P}}$ with entries
\begin{equation}
  \zeta_{i} = \sum_{j=1}^{N_{\text{P}}} w_{ij},
  \label{eq:wei_node_degree}
\end{equation}
is obtained. Here, $\zeta_{i}$ represents all encounters of particle~$i$ with other particles during the observed discrete time interval, regardless of whether they have already met.
However, this vector does not reveal the ``spreader'' particles that enhance macro-mixing most but is biased towards repeated encounters.
Instead, $\boldsymbol{\zeta}$ is later used to normalize the weight matrix $\boldsymbol{W}$ within a spectral clustering approach in order to identify macro-mixing hindering coherent compartments.

Coherent compartments that mitigate global mixing can be identified by solving a normalized cut problem~\cite{shi_normalized_2000}, with the $N_{\text{P}}$ particles forming the vertices (or nodes) and the links represented in the weighted matrix $\boldsymbol{W}$. Let $\boldsymbol{D}\in\mathbb{N}_0^{N_\text{P}\times N_\text{P}}$ be the degree matrix, a diagonal matrix with $d_{\zeta,ii} = \zeta_i$, $i = 1, \ldots, N_{\text{P}}$, where $\boldsymbol{\zeta}$ is the weighted node degree vector from Equation~\ref{eq:wei_node_degree}. The eigenvalue problem
\begin{equation*}
  \boldsymbol{D}^{-1} \boldsymbol{W} \boldsymbol{u} = \lambda \boldsymbol{u}\label{eq:eigenvalue_problem}
\end{equation*}
is equivalent to the generalized eigenvalue problem considered in~\cite{shi_normalized_2000} and forms the basis for the identification of coherent clusters, as explained in the following.

First, note that by construction $\boldsymbol{D}^{-1} \boldsymbol{W}$ is a row-stochastic matrix with real eigenvalues $1 = \lambda_1 \geq \lambda_2 \geq \dots \geq \lambda_{N_{\text{P}}}$ and corresponding real eigenvectors $\boldsymbol{u}_1, \dots, \boldsymbol{u}_{N_{\text{P}}} \in \mathbb{R}^{N_{\text{P}}}$, where $\boldsymbol{u}_1=\mathbf{1}$ is the all-ones-vector. The eigenvectors $\boldsymbol{u}_2, \dots, \boldsymbol{u}_{N_{\text{P}}}$ each have both positive and negative entries. We assume that the network is connected and that $\lambda_1=1$ is a simple eigenvalue. The sign structure of $\boldsymbol{u}_2$ for the eigenvalue $\lambda_2<1$ can then be used to partition the underlying network into two clusters~\cite{shi_normalized_2000}. In our setting, this corresponds to two groups of particles (and thus corresponding fluid volumes) that are closely connected within their group but only have few interactions with the particles of the other group over the respective time span $\Delta t$. That means, when plotting the particle positions in the reactor at a specific time step, while coloring the particles with the corresponding eigenvector entry, the two coherent volumes can be visualized in space. Note that as $\lambda_2$ approaches 1, the two trajectory clusters become increasingly more decoupled.

This idea can be extended to considering more eigenvectors $\boldsymbol{u}_2, \dots, \boldsymbol{u}_k$ to identify $k$ clusters that form coherent compartments. Here $k$ is heuristically chosen based on a spectral gap criterion, i.e.\ $\lambda_{i}-\lambda_{i+1}$ is large for $i=k$ compared to the differences for $i<k$. To extract clusters from several vectors, we use the SEBA (Sparse Eigenbasis Approximation) approach proposed in~\cite{froyland_sparse_2019}.
In that framework the original eigenbasis $\boldsymbol{U} := (\boldsymbol{u}_1, \dots, \boldsymbol{u}_k)$ of the eigenspace $\mathcal{U} \subset \mathbb{R}^{N_{\text{P}}}$ is transformed into a sparse basis $\boldsymbol{B} := (\boldsymbol{b}_1, \dots, \boldsymbol{b}_k)$ of a vector subspace $\mathcal{B} \subset \mathbb{R}^{N_{\text{P}}}$, such that $\mathcal{B} \approx \mathcal{U}$. The matrix $\boldsymbol{B}$ is scaled in such a way that the entry $b_{mn}$ gives the likelihood of the network node (particle, in our case) $m$ belonging to the cluster (i.e.\ coherent compartment) $n$.
By choosing a threshold for this likelihood as detailed in Section~\ref{sec:results}, each particle $m$ can be assigned to a specific cluster $n$ or to the incoherent background, respectively~\cite{froyland_sparse_2019}. The vector $\boldsymbol{s}_{\text{max}} \in \mathbb{R}^{N_{\text{P}}}$ describes an overall cluster probability for each particle with entries $s_{m,\text{max}} = \text{max}_n({b}_{mn})$.

As experimental trajectory data are not uniformly distributed within the reactor, the resulting network can be unconnected for reasonable choices of $\epsilon_r$, with several groups of very few particles forming isolated tiny compartments that do no interact at all with the other fluid volumes. To eliminate these, we only consider the largest connected component of the network, which is extracted by means of SciPy's \texttt{connected\_components}. The eigenvalue problem in Equation~\ref{eq:eigenvalue_problem} is then solved using the sparse matrix function \texttt{eigs} in SciPy for the reduced connected network.

% =========================================================
% =========================================================

\section{Results and discussion}\label{sec:results}
In the following, the results of the 4D-PTV measurement are presented and discussed. Section~\ref{sec:discussion_lagrangian_vel_acc} focuses on the Lagrangian velocities and accelerations.
Section~\ref{sec:discussion_dispersion} analyzes the absolute particle dispersion during the mixing process. From the tracking data, also the Eulerian velocities are calculated by interpolation, and turbulence measures are related to the dispersion analysis, revealing the potential of 4D-PTV tracking also for the measurement of turbulent velocity fields, in Section~\ref{sec:discussion_energy_diss}.
A detailed analysis of the Lagrangian transport and mixing processes via the network-based approach that reveals regions of high mixing and the macro-mixing hindering
compartments is presented in Section~\ref{sec:discussion_network}.

% =========================================================

\subsection{Lagrangian velocities and accelerations}
\label{sec:discussion_lagrangian_vel_acc}

\begin{figure*}[ht!]
\centering
  \includegraphics[width=0.999\linewidth]{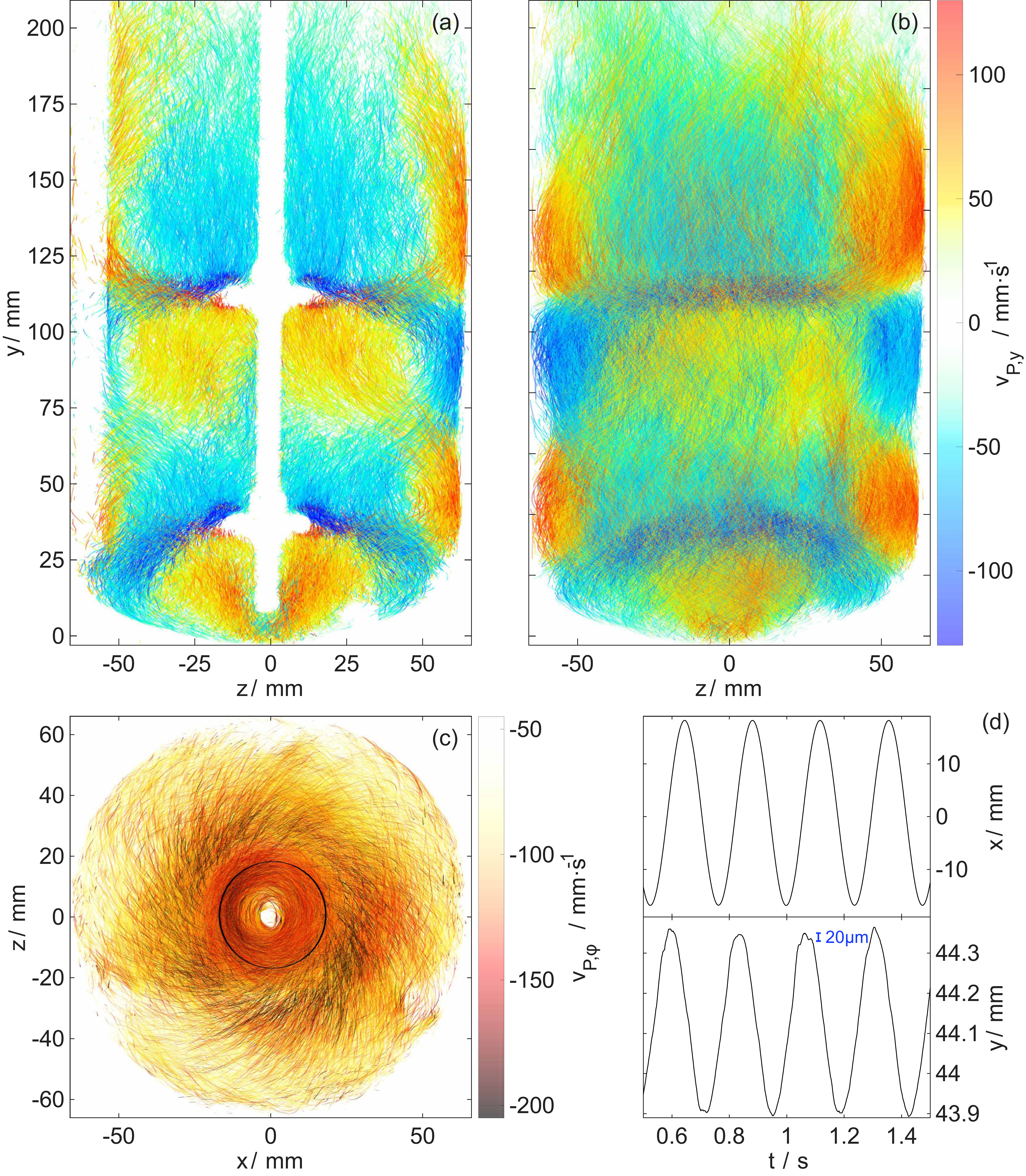}
  \caption{
    Particle trajectories found in the STR colored by their velocity.\@
    (a) \qty{4}{\milli\meter} thick slice along the $y$-$z$ plane through all particle trajectories found during the entire measurement period. The trajectories are colored according to their axial velocity, with those of lower velocity becoming gradually more transparent.\@
    (b) Trajectories shown for the duration of six stirrer rotations ($\Delta t = \qty{0.72}{\second}$) projected to the $y$-$z$ plane.
    The trajectories are colored according to their axial velocity (y-direction), with those of velocities approaching zero being more transparent.\@
    (c) Top-view with trajectories of the length of six stirrer rotations for all particles found within the entire STR.\@
    The trajectories are colored according to their azimuthal velocity, with those of lower velocity and clockwise movement being more transparent.
    The trajectory of a particle attached to the top of the lower stirrer is highlighted in black.\@
    (d) The $x$- and $y$-position are plotted against time for the highlighted particle in (c).
  }
  \label{fig:discussion_track_velocities}
\end{figure*}

Figure~\ref{fig:discussion_track_velocities} shows the particle trajectories found in the STR, colored according to their velocities. The trajectories are drawn slightly transparent, such that a very saturated color represents many overlapping particle trajectories with similar velocities. Furthermore, trajectories of slower particles are more transparent, making the dominant velocities more visible. In Figures~\ref{fig:discussion_track_velocities}a and~\ref{fig:discussion_track_velocities}b, the particle trajectories are colored according to their actual axial velocity, which means that the particles in the red parts of the trajectories move upward, while the particles in the blue parts of the trajectories move downward. In the \qty{4}{\milli\meter} slice through the middle of the reactor shown in Figure~\ref{fig:discussion_track_velocities}a the stirrer shaft, the two Rushton turbines and one baffle on the left-hand side are visible as white voids, which shows that no ghost particles were found there erroneously.
Figure~\ref{fig:discussion_track_velocities}b shows a snapshot of the trajectories of all particles found in the entire STR over the course of six stirrer rotations ($\Delta t = \qty{0.72}{\second}$) projected onto
the $y$-$z$ plane. Therefore, these voids are hidden by particles located in front of and behind the built-in components.

At stirrer heights, a mixing layer with a thickness of around \qty{10}{\milli\meter} is visible in both figures, containing fast upward and downward moving particles.
Furthermore, the velocity coloring shows that particles are drawn into the stirrer from above and below.
At the reactor wall, the outward flow separates into upward and downward streams again.
The separation of particles going up or downward at the heights of the stirrers could tempt one to draw the conclusion that a separation of fluid volumes occurs and separates the STR volume into two compartments that hinder macro-mixing exactly at this height.
However, the more thorough network-based Lagrangian analysis in Section~\ref{sec:discussion_network} will show that the full picture is more intricate and that coherent compartments rather enclose the stirrers.

Figure~\ref{fig:discussion_track_velocities}c shows the particle trajectories colored according to their azimuthal velocity in a top view of the STR.
\@Dark red parts of the trajectories of particles tend to move fast in a clockwise direction, whereas slower clockwise-moving parts are colored light yellow. The few particles moving counterclockwise and those with a very low azimuthal velocity are shown transparently to avoid that particle trajectories at the top of the reactor obscure the dominant velocities at stirrer heights. As expected, the highest azimuthal velocities occur at the stirrer, where the fluid is dragged along and propelled outward by the Rushton turbines. Near the baffles, the azimuthal velocity is close to zero, but the highest axial velocity is found.
Here, Figures~\ref{fig:discussion_track_velocities}a to~\ref{fig:discussion_track_velocities}c also demonstrate that the 4D-PTV measurements presented here cover the entire STR volume with no visible blind spots.
This is crucial for the analysis based on Danckwerts' ideas and implemented as a trajectory network as discussed in Section~\ref{sec:discussion_network}, since the analysis is based directly on Lagrangian trajectories, which cannot be extrapolated in a way similar to a velocity field.

In Figure~\ref{fig:discussion_track_velocities}c, a particle track is highlighted black. This particle was attached to the top side of the lower Rushton turbine during the 4D-PTV measurement and can be used to evaluate the accuracy of the particle motion reconstruction by the calibration and tracking algorithm. Therefore, in Figure~\ref{fig:discussion_track_velocities}d the $x$- and $y$-positions of the particle are plotted against time. Both plots show a sinusoidal oscillation with a period of \qty{0.238}{\second}, which is equal to that of the stirrer. In an ideal experimental setup, the vertical position of a particle sticking to the stirrer would be constant. However, it is impossible to perfectly align all components of an experimental setup, and the stirrer shaft might have been slightly tilted against the reactor walls. Using particles that stick to the Rushon turbine offers the opportunity to calculate the inclination of the stirrer shaft by taking the inverse tangent of the quotient of its $y$-position and $z$-position amplitudes. This gives a tilt angle of approximately \qty{0.7}{\degree}. Also, from the irregularities in $x(t)$ and $y(t)$ from a sine shape function the maximal deviation in position uncertainty of the measurement technique can be derived ($\Delta y_{max} \approx \qty{20}{\um}$, see the blue bar in Figure~\ref{fig:discussion_track_velocities}d), which is smaller than the size of the fluorescent particles themselves. This shows the high precision and repeatability of the position estimation of the particles and thus their trajectories.

\begin{figure*}[ht!]
\centering
  \includegraphics[width=0.87\linewidth]{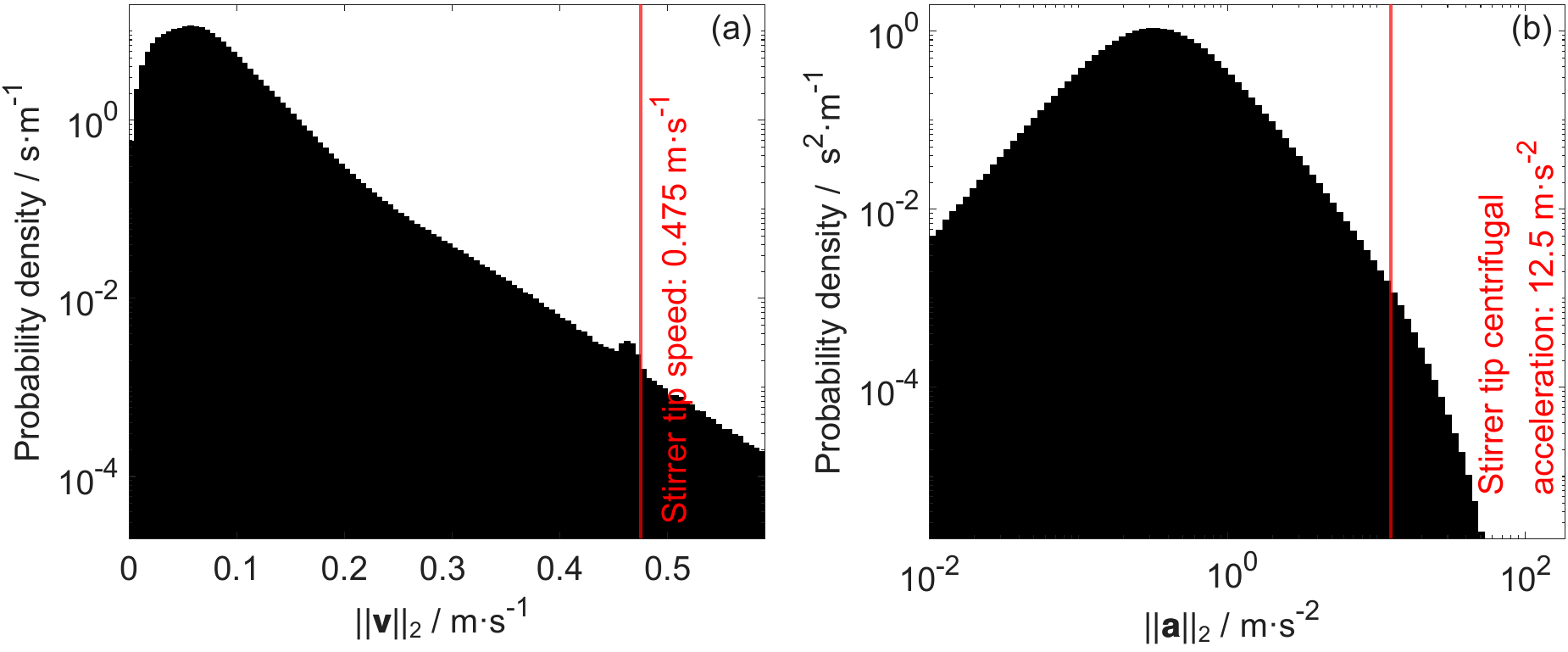}
  \caption{
    Statistical analysis over the entire measurement time for all particles found (88.3 million measurement points).\@(a) Probability density function showing all absolute velocities is presented. The vertical red line represents the stirrer tip speed.\@(b) Probability density function showing all absolute accelerations is presented.}
  \label{fig:discussion_pdf_vel_acc}
\end{figure*}

To create the plots of the probability density function, the absolute Lagrangian velocities of all particles found during the entire measurement are divided into 150 evenly distributed bins ranging from \qty{0}{\m\per\s} to \qty{0.75}{\m\per\s}(Figure~\ref{fig:discussion_pdf_vel_acc}a), while the absolute Lagrangian accelerations are divided into 99 logarithmically spaced bins ranging from \qty[parse-numbers = false]{10^{-2}}{\meter\second\tothe{-2}} to \qty[parse-numbers = false]{10^{2.5}}{\meter\second\tothe{-2}} (Figure~\ref{fig:discussion_pdf_vel_acc}b).
The bin values are divided by the number of particle occurrences of \qty{88.3e6} and the individual bin width to normalize the distributions, so that the area under each curve equals one.
The distribution of absolute particle velocities begins at a probability density value of 0.58 for particles traveling at less than \qty{0.005}{\m\per\s}, increases to a maximum at \qty{0.06}{\m\per\s}, and then decreases to velocities higher than the tip speed of \qty{0.475}{\meter \per \second}.
The latter were previously observed and discussed by Kuschel et al.~\cite{kuschel_validation_2021} using a similar measurement technique but recording only about a third of the STR volume due to unsolved obstruction problems caused by stirrer and baffles.
Notably, such problems have been resolved for the measurements presented here.

From about $\norm{v} \approx \qty{0.2}{\meter\per\second}$ on, the decrease in the velocity probability density function (PDF) is nearly linear, except for a local maximum around the speed of the stirrer tip.
This maximum is partly expected, because fluid directly at the stirrer will have nearly the same velocity as the stirrer, and fluid away from the stirrer will have all possible velocities below the stirrer velocity.
However, some particles sticking to the stirrer might also add to this peak. This general behavior was previously observed for an STR with similar parameters experimentally in a third of the whole volume and in simulations for the entire volume by Hofmann et al.~\cite{hofmann_lagrangian_2022}.

Compared to Hofmann et al., the decreasing zone of the velocity distribution shown here follows a slightly steeper linear trend.
This behavior can be explained by the fact that, due to optical distortion, the bottom of the reactor could not be measured by Hofmann et al. and was therefore disregarded. As the measurement shown here covers the entire reactor volume, including the region at the bottom where the motion is slower, lower velocities are more pronounced in the velocity PDF for our data.
Furthermore, the tracer particles used here are four times smaller in diameter and have a density closer to that of water than those used previously, which makes them better follow the liquid flow. This results in a PDF of the velocity that closely resembles the velocities of the simulated tracers in Hofmann et al.\ more closely than their experimental velocity data.

However, the edges of the velocity PDF should be treated with caution, since particles with a velocity very close to zero are filtered out by the image pre-processing, while particles moving faster than the tip speed of the stirrer are more likely to be lost in particle tracking.
The PDF summarizing the absolute Lagrangian accelerations shows an almost linear upward trend for small accelerations, peaking at around \qty{0.3}{\meter\per\second\squared}. Since many particles move in circles around the $y$-axis of the STR with nearly constant velocity, one might conclude that a peak should occur at zero absolute acceleration.
However, because the particles are forced onto circular paths, they experience at least centripetal acceleration, which shifts the peak of the PDF to the right away from zero. After the peak, the PDF trend decreases linearly until the centrifugal acceleration at the stirrer tip is reached, at which point it drops.
As for the Lagrangian velocity PDF, the plot should be treated with caution for the highest measured accelerations, since the focus of the shown measurement data lies in tracking particles over a long period of time, rather than in capturing the highest occurring accelerations near the stirrers.

\subsection{Particle dispersion}
\label{sec:discussion_dispersion}
\begin{figure*}[ht]
  \centering
  \includegraphics[width=.93\textwidth]{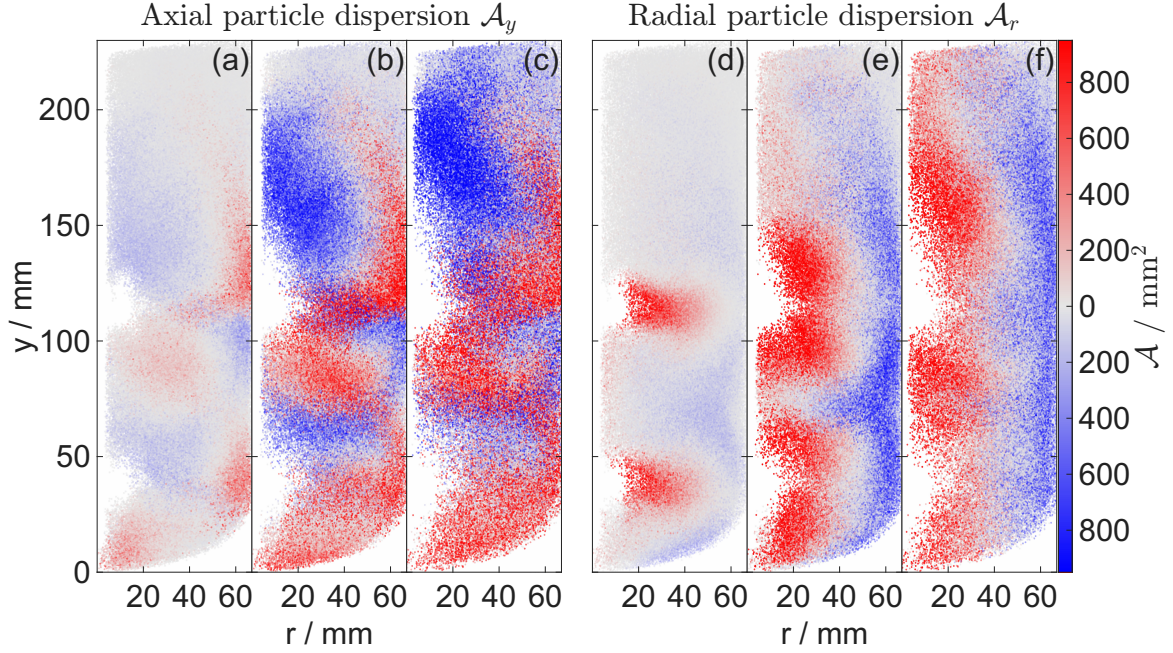}
  \caption{Absolute dispersion of particles in the axial (a, b, c) and  radial (d, e, f) directions after one (a, d), three (b, e) and six (c, f) rotations of the stirrer, plotted against their initial radial and axial positions. Particles moving in the direction of the axes within the respective time interval are colored blue, while particles moving in the opposite direction are colored red.}
  \label{fig:discussion_dispersion_to_position}
\end{figure*}
To gain insight into the mixing behavior of the STR, the classical absolute particle dispersion can be calculated on the full experimental data set. Here, the particle dispersion is defined as the square of the distance a particle travels over a defined time interval.
It is calculated separately for the axial $y$-direction
\begin{equation*}
  \mathcal{A}_{{{P}},y}(t) = \left( {y}_{{{P}}}(t)-{y}_{{{P}}}(t_1) \right )^2
\end{equation*}
and the radial $r$-direction
\begin{equation*}
  \mathcal{A}_{{{P}},r}(t) = \left( {r}_{{{P}}}(t)-{r}_{{{P}}}(t_1) \right )^2 \ \text{with} \ r_{{{P}}} = \sqrt{x_{{{P}}}^2+z_{{{P}}}^2}
\end{equation*}
since the anisotropic reactor geometry causes differences in dispersion in these two directions. Here, $t_1$ denotes the initial time of each particle trajectory.
The dispersion maps shown in Figure~\ref{fig:discussion_dispersion_to_position} are created by summing and plotting all initial positions of the particles in axial and radial space at twelve different initial time steps together, each of which was separated by one stirrer rotation. The particles are colored according to their axial $\mathcal{A}_y$ and radial $\mathcal{A}_r$ dispersion over the following time intervals of one, three, and six stirrer rotations, corresponding to the time intervals $\Delta t = t_{121}-t_{1} = \qty{0.24}{\s}$, $\Delta t = t_{361}-t_{1} = \qty{0.72}{\s}$ and $\Delta t = t_{721}-t_{1} = \qty{1.44}{\s}$. Here, $t_n$ denotes the time of the $n$\textsuperscript{th} point of the particle trajectory.

Figure~\ref{fig:discussion_dispersion_to_position} thus decodes how far each particle traveled in the chosen direction when released from a specific position in the STR.\@
During a single stirrer rotation, minimal particle dispersion is observed in the axial direction but a pattern emerges that becomes more clear for a time interval of three stirrer rotations. A zone of enhanced axial dispersion is clearly seen at stirrer heights. Over the course of six stirrer rotations, this zone widens and becomes more pronounced and reveals well-mixed zones near the impellers, as indicated by upward and downward dispersion within the same region and the simultaneous presence of red and blue colored dots in the same region. The radial particle dispersion for the time interval of one stirrer rotation shows an outward movement near the stirrer tips, which is to be expected given that the Rushton turbine is classified as a radial impeller~\cite{doran_bioprocess_2013,buwa_fluid_2006}.
For the time interval of three and six stirrer rotations, the radial dispersion exhibits the strongest outward movement of particles from the regions above and below the stirrers, from which particles being drawn into the stirrer's working area. However, axial and radial dispersion are both low at the very top of the reactor, indicating zones of very poor mixing. 

\begin{figure*}[!ht]
\centering
  \includegraphics[width=\textwidth]{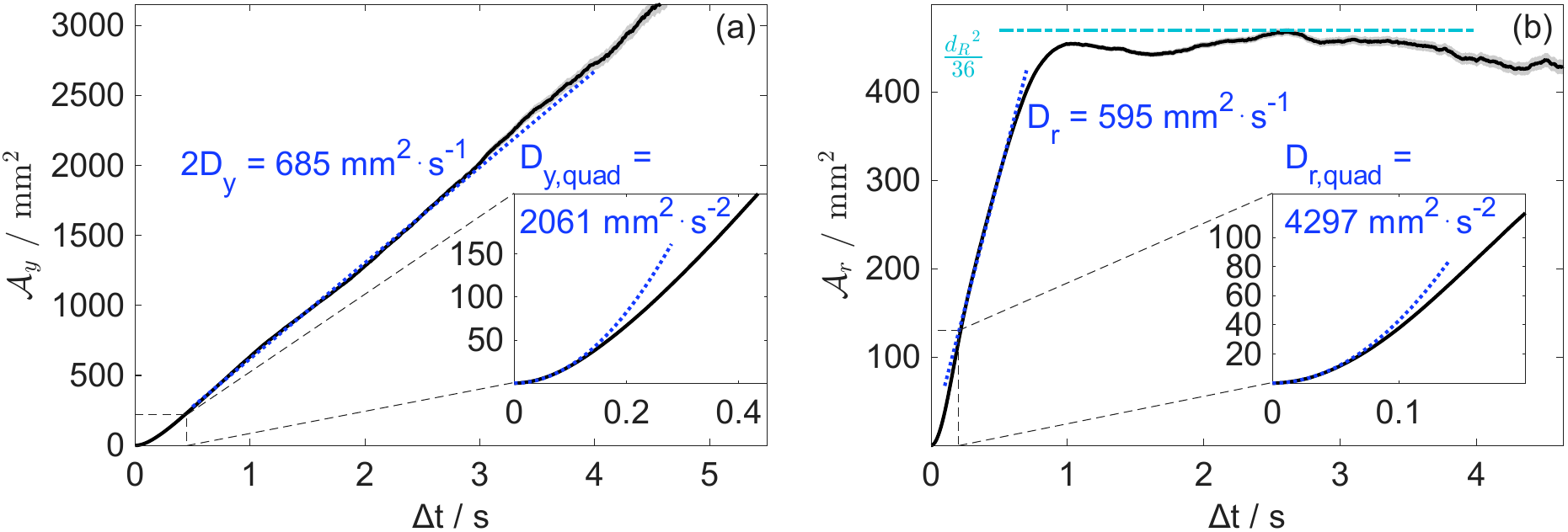}
  \caption{The average axial (a) and radial (b) particle dispersions (black line) and their respective standard errors (gray area) are plotted against time. Dispersion coefficients are determined by fitting a polynomial (blue dashed line) to the data points.}
  \label{fig:discussion_dispersion_vs_time}
\end{figure*}
For a quantification of the overall particle dispersion in the STR the averages of the axial dispersion~$\mathcal{A}_y$ and radial dispersion~$\mathcal{A}_r$ over all particles are plotted versus the observation time $\Delta t = t_n-t_1$ in Figure~\ref{fig:discussion_dispersion_vs_time}. In both directions the dispersion plot follows the ballistic law $\mathcal{A} = D \textsubscript{quad} \cdot (\Delta t)^2$. This behavior is expected since, on small time scales, the velocity vector of a particle does not change with respect to position or time and the consecutive particle velocities are thus highly correlated~\cite{boffetta_transient_1997}.
Therefore, the displacement of the particles is proportional to the observation time. Figure~\ref{fig:discussion_dispersion_vs_time} shows that the time interval during which the dispersion follows the ballistic law is shorter in the radial direction than in the axial direction. The ballistic dispersion coefficients $D_{y,}\textsubscript{quad} = \qty{2016}{\milli\meter\squared\per\second\squared}$ and $D_{r,}\textsubscript{quad} = \qty{4297}{\milli\meter\squared \per \second\squared}$ are obtained by fitting the ballistic law equation to the first \num{51}~time steps for the axial direction and the first \num{21}~time steps for the radial direction, respectively. Both the shorter time interval for ballistic behavior and the higher ballistic dispersion coefficient indicate that mixing occurs faster in the radial direction than in the axial direction. This behavior is consistent with the radial flow impeller classification~\cite{doran_bioprocess_2013,buwa_fluid_2006}.

After a transition zone, the dispersion plot approaches a linear trend~\cite{bakunin_turbulence_2008}. Linear fitting can be used to calculate the dispersion coefficients of $D_y=\qty{343}{\milli\meter\squared\per\second}$ in the axial direction and $D_r=\qty{298}{\milli\meter\squared\per\second}$ in the radial direction.
The nearly identical dispersion coefficients indicate that the particles are dispersing equally in both directions, once their movement is uncorrelated to their initial velocity.
The plot of the radial dispersion flattens at $\mathcal{A}_{r,\text{lim}} \approx \qty{450}{\milli\meter\squared}$ because the particles have a restricted free path length set by the diameter of the reactor.
For a cylindrical geometry with a radius $R_{\text{cyl}}$ the expected value $E_{\text{cyl},r}$ of the radial absolute dispersion after infinite mixing of initially randomly distributed particles can be calculated using the ``Law of the Unconscious Statistician''~\cite{degroot_probability_2012}
\[E_{\text{cyl},r}\! =\!\!\! \int_0^{R_{\text{cyl}}}\!\!\!\!\int_0^{R_{\text{cyl}}}\!\!\!\!\!\!\!\!\!\!\! {d_{\text{diff}}(r_{\text{P}1},r_{\text{P}2})\! \cdot\! f_{\text{cir}}(r_{\text{P}1},r_{\text{P}2})\,\dd r_{\text{P}1}\,\dd r_{\text{P}2}} \]
with
\[d_{\text{diff}} = (r_{\text{P}1}-r_{\text{P}2})^2\]
and the probability density function $f_{\text{cir}}$ for the radial distance of two randomly distributed points in a circle, which here represent the initial and the final radial location of a particle~\cite{ross_probability_2014}:
\[f_{\text{cir}}(r_{\text{P}1},r_{\text{P}2})=f_{\text{cir}}(r_{\text{P}1})\cdot f_{\text{cir}}(r_{\text{P}2})=\frac{2 r_{\text{P}1}}{R_{\text{cyl}}^2} \cdot \frac{2 r_{\text{P}2}}{R_{\text{cyl}}^2} \quad . \]
These equations simplify with $R_{\text{cyl},r} = d_\text{R}/2$ to \[E_{\text{cyl},r} = R_{\text{cyl}}^2/9=d_\text{R}^2/36=\qty{469}{\milli\meter\squared} \gtrapprox \mathcal{A}_{r,\text{lim}} \quad .\]
As the bottom of the STR is curved, restricting the free path length of the particles in the radial direction even further, the limit value is expected to be lower than $E_{\text{cyl},r}$.
In the axial direction, the limit can be approximated in the same way using the probability density function $f_{\text{lin}}(r_{\text{P}1},r_{\text{P}2}) = 1/h_\text{R}^2$, which leads to an expected value of
\[ E_{\text{cyl},y} = \frac{h_\text{R}^2}{6}=\qty{8817}{\milli\meter\squared} \gtrapprox \mathcal{A}_{y,\text{lim}}\quad . \]
This assumes that the probability density does not change with height, which again leads to an overestimation of the dispersion limit $\mathcal{A}_{y,\text{lim}}$ due to the curvature of the vessel's bottom. 
Assuming linear dispersion over time with $D_{y}=\qty{343}{\milli\meter\squared\per\second}$, this limit is reached after \qty{12.9}{\second}. Therefore, the asymptotic behavior cannot be observed in the axial dispersion plot over time shown in Figure~\ref{fig:discussion_dispersion_vs_time}a. To sum up, the Lagrangian data obtained in our measurements can be used to derive precise diffusion coefficients, which can then be used in computational studies or to validate numerical simulations.

\subsection{Eulerian fields}
\label{sec:discussion_energy_diss}
\begin{figure*}[ht!]
\centering
  \includegraphics[width=.90\textwidth]{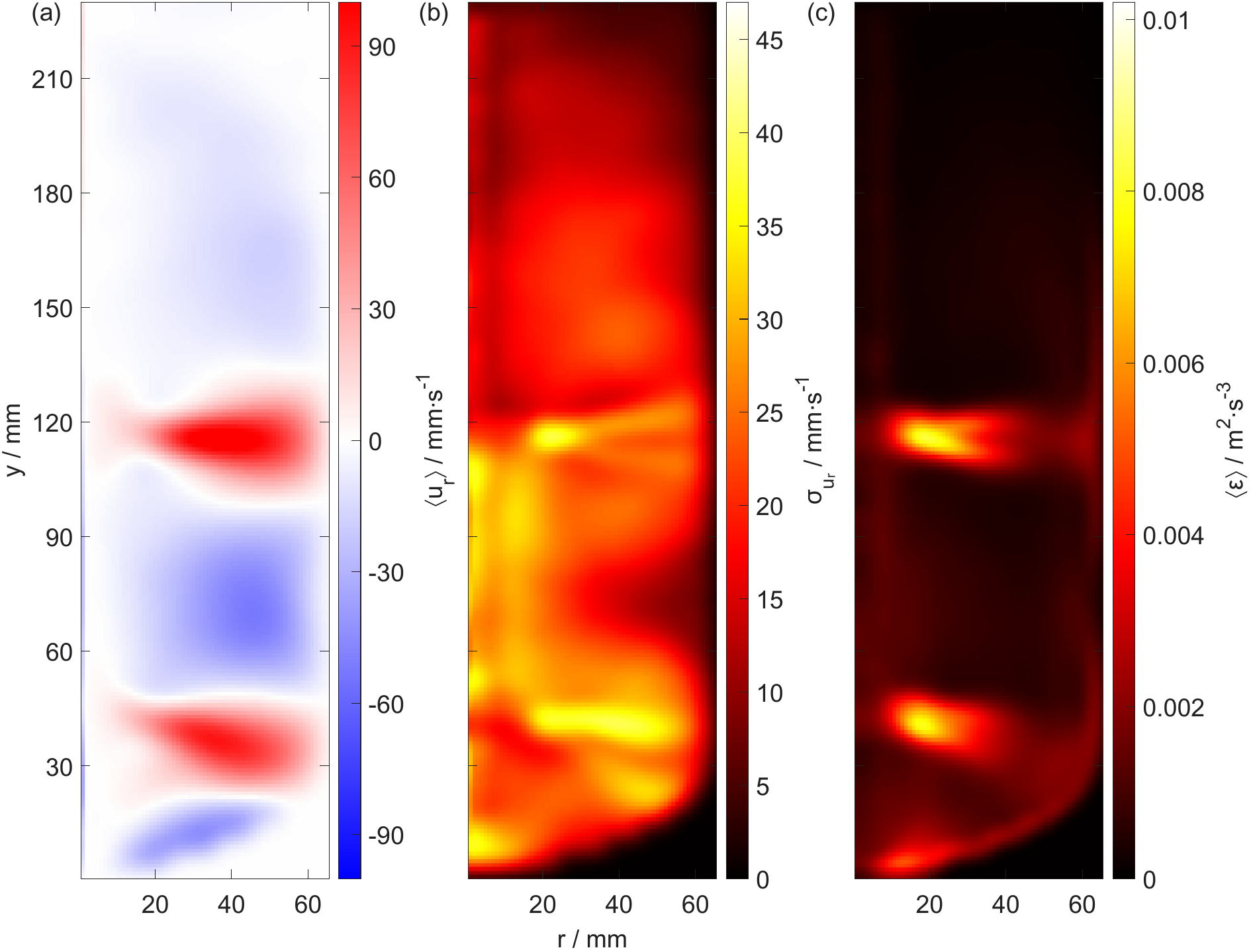}
  \caption{The average~(a) and standard deviation~(b) of the radial Eulerian velocity, and the local energy dissipation rate~(c).}
  \label{fig:discussion_energy_dissipation}
\end{figure*}

Despite the fact that this paper is concerned with the Lagrangian analysis of the mixing within the reactor, the trajectory data are resolved well enough to analyze the time-dependent flow fields for each time step. From these, the spatial gradients of the velocity fields can be computed which are suspected to play a role in the health of microorganisms in bioreactors, especially for mammalian cells~\cite{poertner_review_2024}. With the data presented, it is therefore also possible to estimate the stresses that a microorganism experiences on its journey through the reactor, the so called lifelines~\cite{hofmann_lagrangian_2025, kuschel_simulated_2020, haringa_analysis_2023}. Also, from the instantaneous velocity fields, the mean velocity fields and mean energy dissipation also can be computed.
Figure~\ref{fig:discussion_energy_dissipation}a shows the mean radial velocity, averaged azimuthal and over time, evaluated on an Eulerian grid, as described in Section~\ref{sec:processing}. At stirrer height, the fluid is propelled outward, while above and below the stirrers the fluid is sucked into the center of the reactor, resembling the behavior of the radial dispersion for short times in Figure~\ref{fig:discussion_dispersion_to_position}d. 
Again, this is to be expected because the Rushton impellers are classified as radial impellers. The average radial velocity is used to calculate the standard deviation of the radial velocity $\sigma_{u_r}$ at each radial position against height such that $v_r \pm \sigma_{v_r}$ statistically covers \qty{68.3}{\percent} of the existing radial velocities at each location.
The standard deviation, shown in Figure~\ref{fig:discussion_energy_dissipation}b, indicates how much the flow fluctuates over time at each location. The radial velocity fluctuates widely above and below the stirrer, as can be concluded from the high values of $\sigma_{v_r}$ (yellow). According to turbulence theory, a high fluctuating velocity is also associated with a larger tracer dispersion~\cite{bakunin_turbulence_2008}, which complies with the findings for the absolute dispersion here; see Figures~\ref{fig:discussion_dispersion_to_position}e,f.

\begin{figure*}[ht!]
  \centering
  \captionsetup[subfigure]{skip=-260pt,slc=off,margin=0pt}
  \begin{subfigure}{.45\textwidth}
    \includegraphics[width=\textwidth]{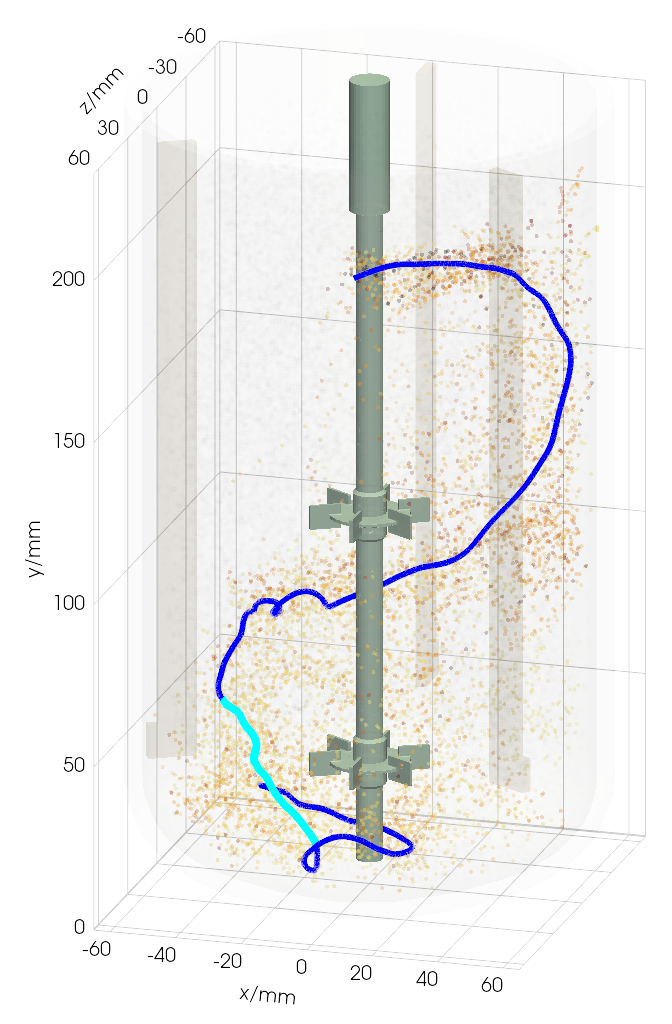}
    \caption{}
    \begin{tikzpicture}[remember picture, overlay]
      \draw[->, black, thick] (1.2,-4) to [bend left=45] (2.5,-2.5) to [bend right=45] (3, -1.8) to [bend right=45] (2.5, -1.5);
    \end{tikzpicture}
  \end{subfigure}
  \begin{subfigure}{.45\textwidth}
    \includegraphics[width=\textwidth]{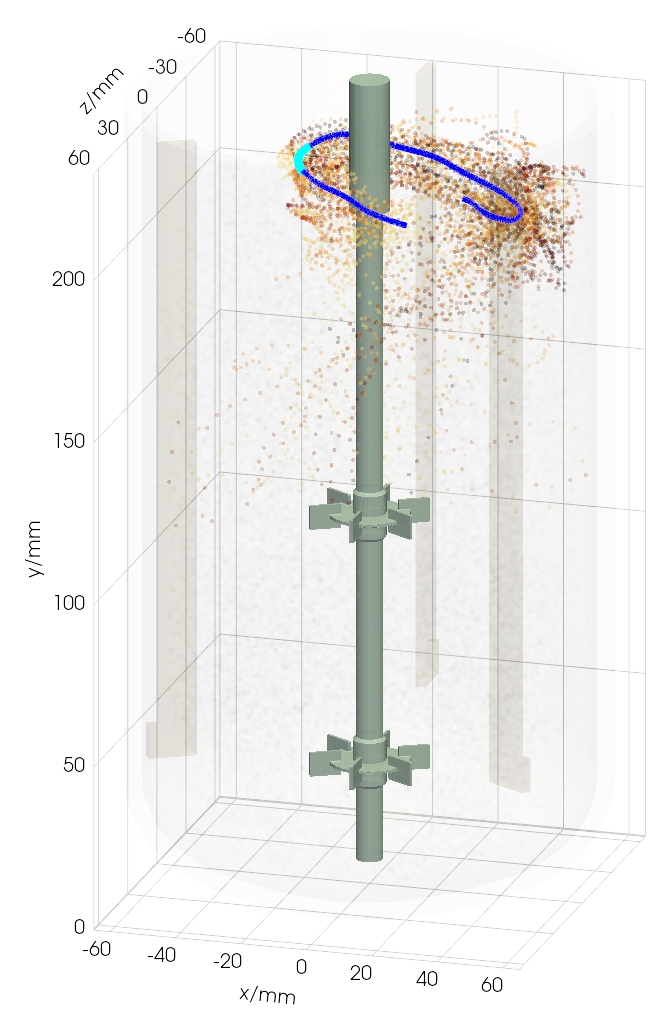}
    \caption{}
    \begin{tikzpicture}[remember picture, overlay]
      \draw[->, black, thick] (2.5,0.8) to [bend left=45] (2.3,1.5) to [bend left=45] (3.3,1.7);
    \end{tikzpicture}
  \end{subfigure}
  \vspace{240pt}
  \begin{subfigure}{.087\textwidth}
    \includegraphics[width=\textwidth]{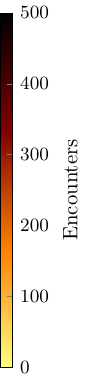}
    \vspace{-210pt}
  \end{subfigure}
  \caption{The positions of the particles that contribute to the node degree of the particle following the blue colored trajectory are plotted at every \num{50}\textsuperscript{th} time step.
  The particles are colored by the number of encounters with the particle following blue trajectory (weighted node degree, see Equation~\ref{eq:wei_node_degree}).
  Two different particles are depicted. Within the same time interval one particle $\text{P}_{\text{spread}}$, shown in (a), is moving from the bottom to the top, while the other particle $\text{P}_{\text{local}}$, shown in (b), is remaining in the upper part of the reactor. For reference, the light blue section of both tracers represent three stirrer rotations (time span is $t \in [\qty{1.20}{\s}, \qty{1.92}{\s}]$).
  Arrows indicate movement direction of each tracer.
  }
  \label{fig:discussion_encounter}
\end{figure*}
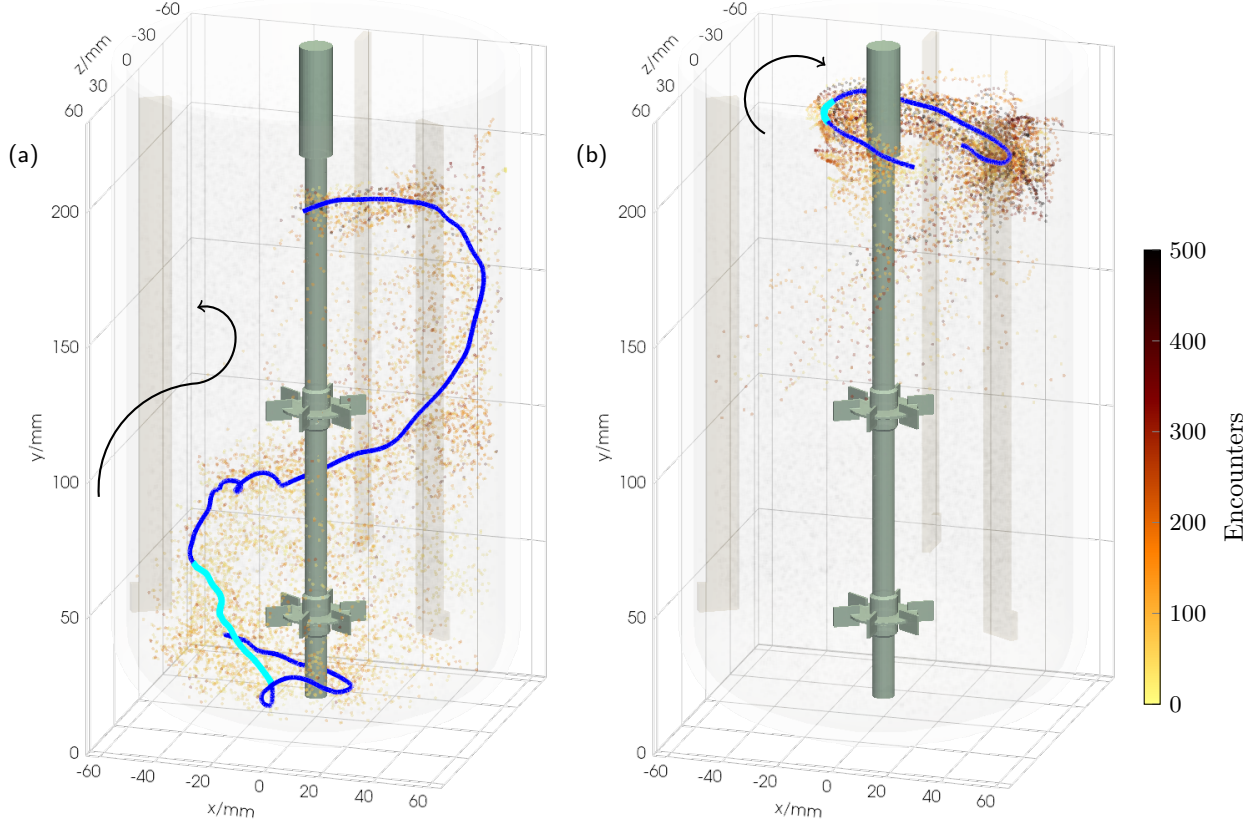

Figure~\ref{fig:discussion_energy_dissipation}c shows the local mean energy dissipation, calculated as detailed in Section~\ref{sec:processing}, again averaged over time and radius.
The figure shows areas of high mean energy dissipation close to the stirrers.
As for the standard deviation, but less pronounced, these areas divide into an upper and a lower segment radiating from each stirrer tip outwards.
This indicates that most of the input energy is dissipated near the top and bottom of the stirrer, where the high speed fluid pushed by the stirrer meets the slow flow above and below the stirrers which causes high spatial velocity gradients.
This plot also aligns well with the radial dispersion shown in Figure~\ref{fig:discussion_dispersion_to_position}d, supporting the assertion that faster mixing is correlated with higher local energy input.
The knowledge of these Eulerian fields derived from the Lagrangian trajectory data thus allows us to analyze these inhomogeneities in order to estimate the possible effects on microorganisms. 
However, Eulerian fields do not provide the information about volumes that mix poorly with their surroundings. 
In contrast, the mean energy dissipation and the standard deviation of the radial velocity both suggest that the areas around the stirrers are well mixed. 
The subsequent Lagrangian analysis in the next section reveals that this is really the case; however, it also reveals much more: The spatial extension of the well-mixed coherent compartments and the separation between them, which act as mixing barriers.

\subsection{Mixing analysis based on node degree}
\label{sec:discussion_network}
\begin{figure*}[ht]
  \centering
  \includegraphics[width=0.44\textwidth]{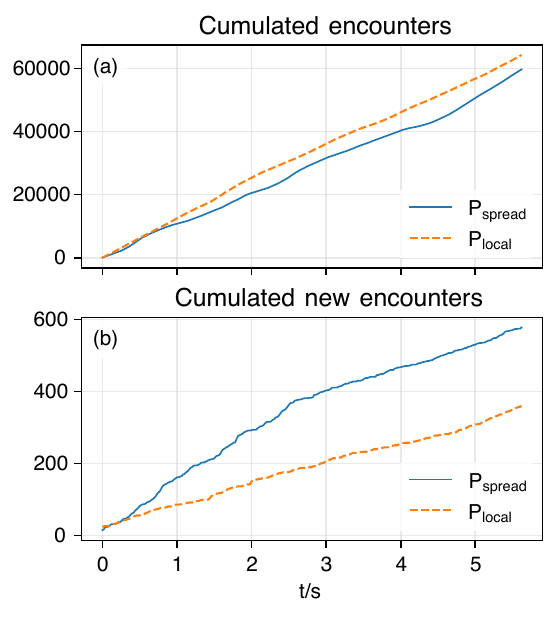}
  \includegraphics[width=0.44\textwidth]{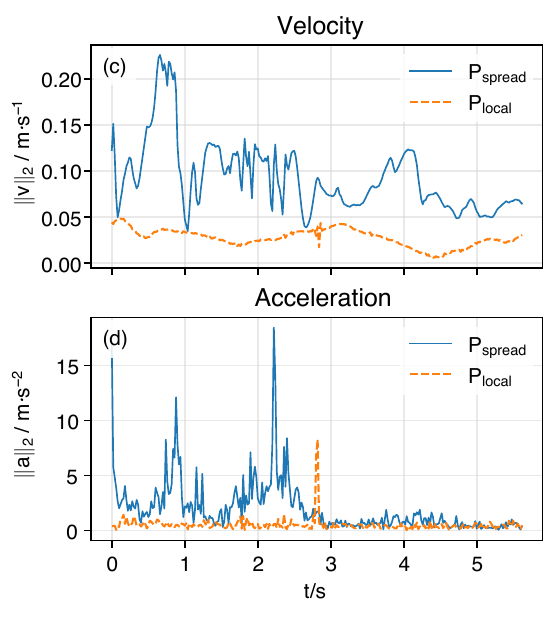}
  \caption{Cumulated encounters (a), new encounters (b) of both trajectories from Figure~\ref{fig:discussion_encounter} and the velocity (c) and acceleration (d) along the trajectories. The cumulated encounters curves reflect the increasing weighted node degree as more time slices are considered, while the cumulated new encounters curves do so for the unweighted node degree.}
  \label{fig:discussion_encounter_vs_time_vel_acc}
\end{figure*}

The trajectory-based network analysis was conducted as described in Section~\ref{sec:methods_network}. In total, approximately \num{217000}~trajectories were considered. The main parameter in the network construction is the encounter radius $\epsilon_r$, which has to be chosen such that the resulting network is connected (or only has very few isolated groups of nodes that can be eliminated) but sparse.
Here we choose $\epsilon_r=7$ mm $\approx 3 d_{\text{avg}}$, where $d_{\text{avg}}=2.3\pm 0.02$ mm is the average distance between particles and their nearest neighbor, with the latter being the standard deviation.
However, the following results are insensitive to the choice of $\epsilon_r$ as has been shown in previous studies for other Lagrangian data from geophysical flow~\cite{padberg-gehle_networkbased_2017}, in the sense that increasing $\epsilon_r$ will lead to very similar results.

To illustrate the possibilities that the Lagrangian trajectories provide to perform the analysis envisioned by Danckwerts, two trajectories with a very different meaning with regard to mixing are depicted in Figure~\ref{fig:discussion_encounter}, colored blue.
The surrounding particles are colored according to the number of encounters with the corresponding blue trajectory, and their positions are indicated at each \num{50}\textsuperscript{th} time step.
Particles without such encounters are not shown. The blue particle trajectories in Figures~\ref{fig:discussion_encounter}a and~\ref{fig:discussion_encounter}b have very different overall displacements, but the same temporal duration of $t_\text{total} = \qty{5.6}{s}$, which is approximately equivalent to \num{23.5}~stirrer rotations.
While particle $\text{P}_{\text{spread}}$, shown in Figure~\ref{fig:discussion_encounter}a, travels through the entire reactor, has a trajectory length of approximately $\sum_{k=1}^{N_T}\norm{\boldsymbol{x}_i(t_k)-\boldsymbol{x}_i(t_{k-1})} =\qty{523.2}{\mm}$, and encounters many different particles only a few times, particle $\text{P}_{\text{local}}$, shown in Figure~\ref{fig:discussion_encounter}b, remains in the upper part of the reactor, has a much shorter trajectory length of approximately \qty{155.6}{\mm}, and repeatedly encounters the same particles.
Figures~\ref{fig:discussion_encounter_vs_time_vel_acc}a and~b illustrate this, plotting the cumulative numbers of all encounters and of encounters with new particles against the travel time of $\text{P}_{\text{spread}}$ and $\text{P}_{\text{local}}$, respectively.
In Figure~\ref{fig:discussion_encounter_vs_time_vel_acc}a the cumulative encounters demonstrate linear behavior, with a similar increase for both particles. This is to be expected since the increase only depends on the local particle seeding density and the encounter radius $\epsilon_r$.
What we can conclude from this plot is that the seeding density is very uniform along two randomly picked trajectories at different heights. In Figure~\ref{fig:discussion_encounter_vs_time_vel_acc}b, only encounters with particles that have not been met before are considered.
The plot for $\text{P}_{\text{spread}}$ shows a steeper incline than the plot for $\text{P}_{\text{local}}$, indicating that $\text{P}_{\text{spread}}$ encounters more new particles per time step than $\text{P}_{\text{local}}$. As $\text{P}_{\text{spread}}$ reaches the upper part of the STR, and therefore shares the same region with $\text{P}_{\text{local}}$, the two curves run in parallel.

\begin{figure*}[ht]
  \centering
  \includegraphics[width=.96\linewidth]{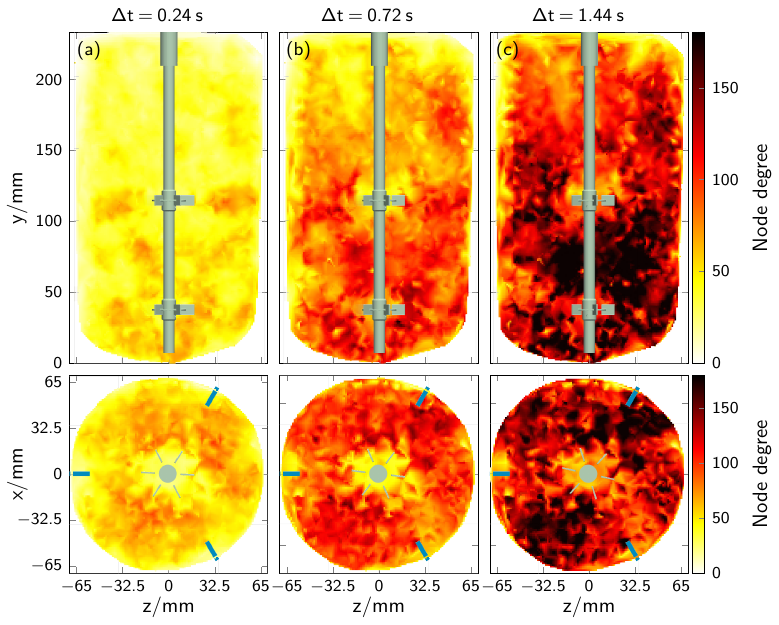}
 \caption{Unweighted node degree at $t = \qty{1.56}{\s}$, evaluated for a time range (a) $t \in [\qty{1.44}{\s}, \qty{1.68}{\s}]$, (b) $t \in [\qty{1.20}{\s}, \qty{1.92}{\s}]$ and (c) $t \in [\qty{0.84}{\s}, \qty{2.28}{\s}]$. The integration time is equivalent to 1, 3 and 6 stirrer revolutions respectively for a stirrer frequency of \qty{252}{rpm}. Upper row view at $x = 0$ and lower row view at $y = \qty{112}{\milli\meter}$.}
  \label{fig:discussion_node_degree_u}
\end{figure*}
Both particles therefore have a high weighted node degree but very different contributions to mixing. $\text{P}_{\text{spread}}$ has a noticeably larger unweighted node degree than $\text{P}_{\text{local}}$.
If $\text{P}_{\text{spread}}$ is viewed as a fluid parcel, it can be attributed to be a ``spreader'' that transfers its reactants to many other fluid parcels across the whole reactor and thus enhances macromixing.
In contrast, the fluid parcel visualized by $\text{P}_{\text{local}}$ will quickly reach the same reactant concentration as the permanent neighbors that travel with it through the reactor and is thus only important for more local meso- and micromixing.
In fact, as described in Section~\ref{sec:methods_network}, this particle and its neighbors might well be a part of a coherent compartment that hinders macromixing.

Figures~\ref{fig:discussion_encounter_vs_time_vel_acc}c and~d show the velocity and acceleration along the two depicted trajectories plotted against time. It can be seen that $\text{P}_{\text{spread}}$ experiences significantly higher and more frequent peaks in velocity and acceleration compared to $\text{P}_{\text{local}}$. If we consider the particles to be cells in a bioreactor, we can deduce important information for bio processes from the time-resolved particle tracks. For instance, microorganisms can alter their metabolism in response to velocity gradients~\cite{jones_continuous_2019,simoes-faria_wall_2025}.
Therefore, if two microorganisms follow the trajectories of $\text{P}_{\text{spread}}$ and $\text{P}_{\text{local}}$, it is likely that their metabolisms differ from each other, leading to inhomogeneities in product concentration in the STR and resulting in a change in overall production and yield, given the potential abundance of different paths and therefore different metabolisms.

\begin{figure*}[ht!]
  \centering
  \includegraphics[width=\linewidth,page=1]{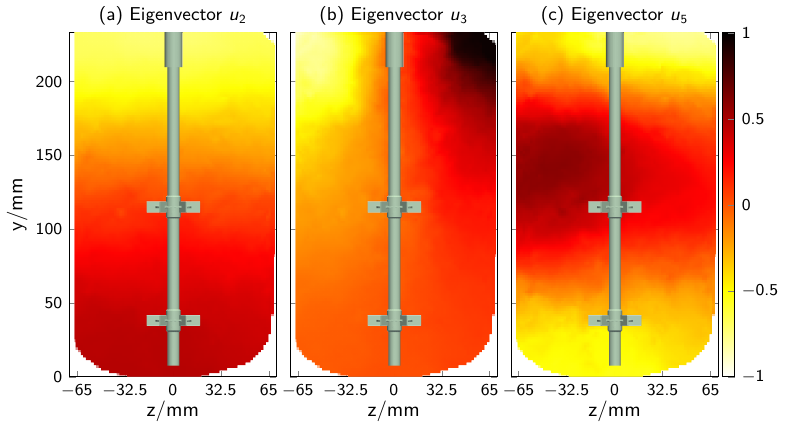}
  \caption{Leading eigenvectors of $\boldsymbol{D}_\zeta^{-1} \boldsymbol{W}$ highlight compartments for three stirrer rotations $t \in [\qty{1.20}{\s}, \qty{1.92}{\s}]$, plotted at $t = \qty{1.56}{\s}$. View at plane $x = 0$.}
  \label{fig:discussion_eigenvals}
\end{figure*}
To gain insight into the flow behavior within the entire STR, the exemplary node degree analysis is extended to all particles.
For the following studies, time periods equivalent to one ($\Delta t = \qty{0.24}{\s}$), three ($\Delta t = \qty{0.72}{\s}$) and six ($\Delta t = \qty{1.44}{\s}$) stirrer rotations were considered as for the absolute dispersion in Section~\ref{sec:discussion_dispersion}.
The unweighted node degree $\boldsymbol{\xi}$ from Equation~\ref{eq:unw_node_degree} 
is evaluated over these time intervals and plotted in Figure~\ref{fig:discussion_node_degree_u} at the particle locations at time $t = \qty{1.56}{\s}$, the center of all three time intervals. 

The highest unweighted node degree values can be found around the stirrer, indicating strong mixing for all three stirrer time spans evaluated here.
There are particularly low values of the unweighted node degree in the upper edge region of the vessel in the positive $y$-direction.
These indicate, that particles there generally meet less other particles, even though the particle seeding density is comparable.

Finally, the node degree at the height of the upper stirrer of $y = \qty{112}{\milli\meter}$ is shown in the lower row of the Figure~\ref{fig:discussion_node_degree_u}. As is evident, there is a higher prevalence of high node degree values in the vicinity of the stirrer, for the shortest time interval, suggesting a significant mixing intensity in this region.

% =========================================================

\subsection{Lagrangian compartments and transport analysis}

\begin{figure*}[ht!]
  \centering
  \includegraphics[page=2]{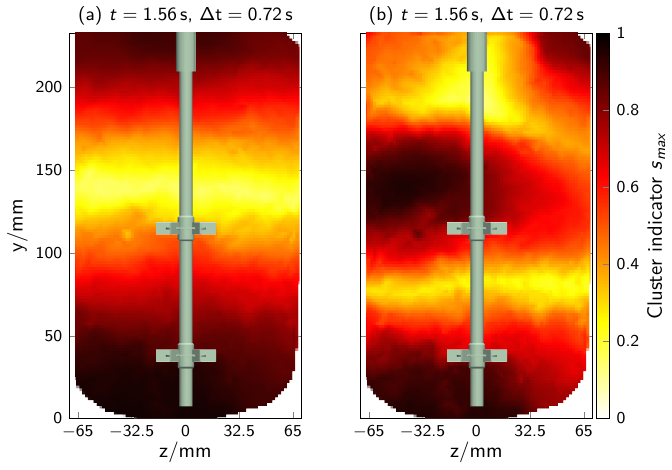}
  \caption{Identification of coherent compartments via the sparse eigenbasis approximation SEBA for three stirrer rotations $t \in [\qty{1.20}{\s}, \qty{1.92}{\s}]$.
  (a) two coherent sets, located at the top and the bottom of the reactor, are obtained from the first two eigenvectors.
  (b) SEBA applied to the first five eigenvectors gives a more refined picture of the mixing compartments.
  View at plane $x = 0$.
  }
  \label{fig:discussion_smax3rot}
\end{figure*}
 The information how much a single particle is involved in a coherent compartment, where the particles primarily meet only other particles from the same coherent compartment, is contained in the network matrix $\boldsymbol{W}$.
 This information can be evaluated by solving the spectral clustering problem based on the eigenvalue value problem formulated in Equation~\ref{eq:eigenvalue_problem}, where the identified clusters form the coherent compartments.
 As regards the structure of the network, the presence or absence of links ($w_{ij} > 0$ or $w_{ij} = 0$) in groups of trajectories is the strongest indicator of clustering.
 Furthermore, the weights on the links (which provide information about the number of encounters or interaction intensity) also influence the structures of the clusters.

More to the point, the normalized network weight matrix $\boldsymbol{D}_\zeta^{-1} \boldsymbol{W}$ in Equation~\ref{eq:eigenvalue_problem} used for clustering, contains information of the encounters of each particle in relation to all its encounters, since every entry of the network weight matrix $\boldsymbol{W}$ is normalized by the overall contacts of each particle using the degree matrix $\boldsymbol{D}_\zeta$.

This can be exemplarily be illustrated using Figure~\ref{fig:discussion_encounter}.
The non-zero entries in the $i^\text{th}$ row of $\boldsymbol{D}_\zeta^{-1} \boldsymbol{W}$ are larger for the $i^\text{th}$ particle being the blue particle $\text{P}_{\text{local}}$ in Figure~\ref{fig:discussion_encounter}b than for the $i^\text{th}$ particle being $\text{P}_{\text{spread}}$ in Figure~\ref{fig:discussion_encounter}a.
The reason is that for the particle $\text{P}_{\text{spread}}$ the corresponding row in $\boldsymbol{W}$ has many non-zero entries due to the many different contacts and by row-normalizing with the weighted node degree these get very small in magnitude.
For $\text{P}_{\text{local}}$, in contrast, the corresponding rows in $\boldsymbol{W}$ and $\boldsymbol{D}_\zeta^{-1} \boldsymbol{W}$ have fewer but larger non-zero entries because of repeated encounters with the same particles. Note that while the trajectories for $\text{P}_{\text{spread}}$ and $\text{P}_{\text{local}}$ shown in Figure~\ref{fig:discussion_encounter} cover a long time span of approximately \num{23.5}~stirrer rotations, the following identification of compartments will only use the Lagrangian information of three stirrer rotations. Still, $\text{P}_{\text{local}}$ seems highly likely to be part of a compartment at the top of the reactor, whereas $\text{P}_{\text{spread}}$ wanders between different compartments as will be discussed in the following.

Returning to the spectral clustering problem, the leading eigenvectors of $\boldsymbol{D}_\zeta^{-1} \boldsymbol{W}$ obtained from trajectories for three stirrer rotations are shown in Figure~\ref{fig:discussion_eigenvals}. For better comparison with the node degree results, the eigenvector data, which are available for the whole three-dimensional vessel, are plotted in a slice to the $x = 0$ plane. In particular, $\boldsymbol{u}_2$ highlights regions with different connectivities at the top and bottom of the vessel, while $\boldsymbol{u}_3$ and $\boldsymbol{u}_4$ (not shown) provide a finer partition of the top region. A central connected region is highlighted by $\boldsymbol{u}_5$.

To identify coherent compartments from these eigenvectors, the sparse eigenbasis approximation approach SEBA from~\cite{froyland_sparse_2019} is used and a cluster indicator $\boldsymbol{s}_{\text{max}}$ is obtained.
The result of such post-processing of $\boldsymbol{u}_1$ and $\boldsymbol{u}_2$ is shown in Figure~\ref{fig:discussion_smax3rot}a, where again the three-dimensional results are limited to the $x = 0$ plane.
In this case, two coherent areas are identified, one at the bottom and one at the top of the vessel, indicated by high values of $\boldsymbol{s}_{\text{max}}$.
Five coherent sets are obtained when $\boldsymbol{u}_1, \dots, \boldsymbol{u}_5$ are considered, see Figure~\ref{fig:discussion_smax3rot}b.
In this plot, only four of them are visible due to the two-dimensional presentation and the orientation of the plane. In particular, a coherent compartment can be identified in the center of the vessel.
Intriguingly, the two coherent sets in the center and bottom of the vessel are characterized by strong internal mixing, which is also evident in both the node degree study in Figure~\ref{fig:discussion_node_degree_u} and the absolute dispersion in Figure~\ref{fig:discussion_dispersion_to_position}.

\begin{figure*}[ht!]
  \centering
  \captionsetup[subfigure]{skip=-280pt,slc=off,margin=0pt}
  \begin{subfigure}{.49\linewidth}
    \includegraphics[width=\linewidth]{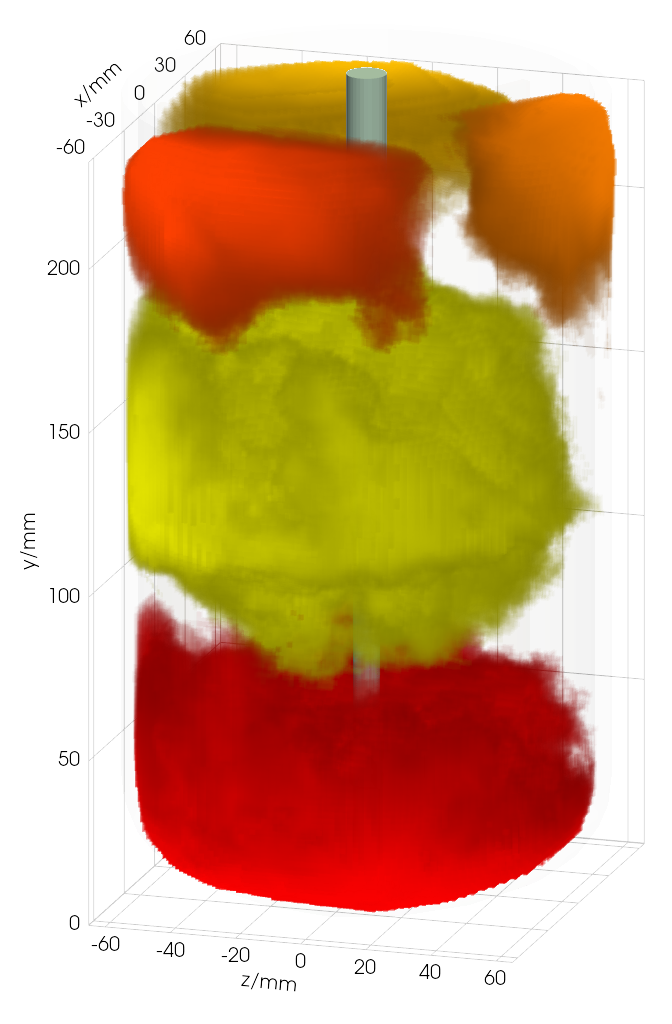}
    \caption{}
  \end{subfigure}
  \begin{subfigure}{.49\linewidth}
    \includegraphics[width=\linewidth]{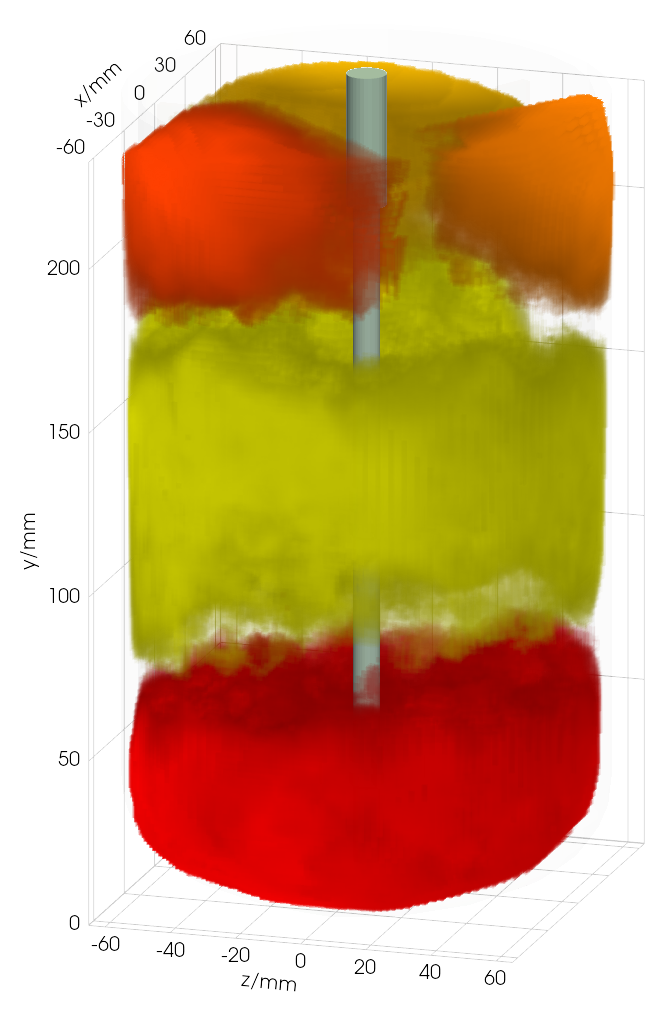}
    \caption{}
  \end{subfigure}
  \vspace{268pt}
  \caption{Extraction of five coherent sets: Only those particles are volumized for which the $\boldsymbol{s}_{\text{max}}$ cluster indicator exceeds $0.5$. (a) Particle positions at $t_1 = \qty{1.20}{\s}$. (b) Particle positions at $t_{N_T} = \qty{1.92}{\s}$. While there appears to be strong mixing within the detected compartments, there seems to be some transport barrier between the lower and central compartment in the observed span of three stirrer rotations ($\Delta t = \qty{0.72}{\s}$).}
  \label{fig:discussion_clusterc5}
\end{figure*}
Note that the high node degree values occur in the middle of the two lower compartments and do not coincide with the boundaries of these compartments as one might expect.
This indicates that these two compartments are well-mixed within and that the area with a high node degree is not a strict ridge dividing the transport in the reactor into different dynamical systems. Instead, the areas directly above and below each stirrer appear to form one coherent compartment.
Meanwhile, the Lagrangian mixing between the agitation stages (between the stirrers) is greatly reduced. These results align well with those of experiments involving electrolyte pulse injections~\cite{alves_alternative_1997}.

In contrast, the three upper compartments appear to be dead zones with little internal mixing.
However, their existence and dynamics are of great importance for bioreactors because substrate feed is most often injected from above for hygienic reasons~\cite[p.~946]{liu_bioprocess_2013}.
If a feed occurs into one of these compartments it would greatly hinder the initial macro- and mesomixing.

We extract coherent compartments as volumes by means of setting a threshold to the cluster indicator $\boldsymbol{s}_{\text{max}}$, taking into account only particles that exceed this threshold.
The resulting compartments for the case of five clusters and a threshold of $0.5$ are shown in Figure~\ref{fig:discussion_clusterc5}, where Figure~\ref{fig:discussion_clusterc5}a shows particles that make up coherent compartments at the beginning of the time interval (i.e.\ at $t_1 = \qty{1.20}{\s}$) and Figure~\ref{fig:discussion_clusterc5}b shows the same compartments after three stirrer rotations (i.e.\ at $t_{N_T} = \qty{1.92}{\s}$), showing how the coherent compartments evolve over time.

Altogether, the findings achieved by the trajectory-based network indicate five compartments: one around the lower stirrer, one around the upper stirrer, and three in the upper region of the vessel, each of which is located between the baffles. The transport between compartments seems to be very low, whereas the interaction of particles within each of the two larger compartments is high as analyzed by the node degree.
Comparing these results with the long-time trajectories of $\text{P}_{\text{local}}$ and $\text{P}_{\text{spread}}$ in Figure~\ref{fig:discussion_encounter}, we observe that the particle $\text{P}_{\text{local}}$ is confined to one of the slowly rotating top compartments, whereas $\text{P}_{\text{spread}}$ starts in the bottom compartment and stays there for a considerable time span before it moves to the top via the central compartment.
The clustering and node degree results agree very well with the former simulation results in~\cite{weiland_computational_2023}, where finite-time Lyapunov exponents have been used to analyze regions of strong mixing. The advantage of the network-based approach using the node degree and spectral clustering for the quantification of mixing and the identification of coherent compartments used here is that the structures that dominate macromixing can be derived. Further, in contrast to the finite-time Lyapunov exponent analysis, the network-based approach can be directly applied to experimentally measured trajectory data as it does not require the tracer particles to be started on an equidistant grid. 

% =========================================================
% =========================================================

\section{Conclusion}\label{sec:conclusion}
This work analyzes the mixing behavior in a chemical reactor, following early scientific analysis ideas of tracking the reactants during their entire journey through the vessel by Danckwerts and Levenspiel.
The reactor of this study is a laboratory-scaled \qty{3}{\liter} STR stirred by two Rushton turbines; one of the most commonly used reactors in process engineering and chemistry, especially in the early process development stages.
The time-resolved flow in the entire volume of the vessel was experimentally measured using 4D-PTV, following approximately \num{40000} tracer particles in every time step with a spatial position reconstruction accuracy of \qty{20}{\micro\meter}.

The mass transport within the STR is characterized by analyzing the Lagrangian motions of the particles. The Lagrangian particle movement reveals fast outward streams at stirrer heights that separates at the vessel wall.
Also at stirrer heights, a mixing layer with a thickness of around \qty{10}{\milli\meter} extends from the stirrers to the wall.
Lagrangian statistics of the measured data summarizing all velocities and accelerations found during the entire measurement show a high level of consistency with previous publications on numerical data.
The absolute dispersion over time follows ballistic behavior at short timescales, linear behavior at intermediate timescales, and reaches saturation due to geometric constraints for long times as expected from theoretical considerations.
Absolute dispersion coefficients in axial and radial direction are derived from the data that can be used, e.g., for compartment modeling studies numerical simulation validation.
Field variables such as the locally resolved energy dissipation are also obtained from a Eulerian evaluation of the velocity data.
The energy dissipation and the standard deviation of the radial velocity show maxima directly above and below the stirrer tips which complies with the areas of largest particle dispersion.
Network-based Lagrangian mixing analysis methods show zones of strong and poor mixing.
Spectral clustering reveals five coherent compartments with low mass transfer between them but partly with strong internal mixing.
These coherent compartments can largely hinder macromixing, and their specification is thus important to characterize mixing dynamics in biochemical and chemical reactors.
The derived network methods can be valuable to support scale up processes and develop feeding strategies to mitigate substrate inhomogeneities in biochemical reactors.

Overall, this study demonstrates the effectiveness of using 4D-PTV measurements combined with network-based Lagrangian analysis to identify coherent structures, transport barriers and areas of high mixing in complex flows.
From a combination of the Lagrangian trajectories and the Eulerian field derivatives, the experienced forces and stresses of a passive tracer can now also be evaluated, providing new possibilities for insights into the flow conditions experienced by microorganisms.
Further, the data presented here is currently used for the development of Lagrangian analysis methods that incorporate diffusion and reactions of the fluid parcel along its trajectory through the vessel as formerly envisioned by Danckwerts.
Also, the data set, which is made freely available alongside this publication can serve for the validation of computational simulations.
In future measurements the details of the high velocity gradients near the stirrer tips and reactors with gas hold-up will be considered to broaden the applicability of the generated insights.

% =========================================================
% =========================================================

\section*{Acknowledgements}
This project is funded by the Deutsche Forschungsgemeinschaft (DFG, German Research
Foundation) -- 503850735 -- CRC 1615 SMART reactors under usage of major research instrumentation MUST at the Hamburg University of Applied Science funded by the DFG -- 514139948
\newpage

\bibliographystyle{elsarticle-num}
\bibliography{references}
\end{document}